\documentclass[usenatbib,onecolumn]{mn2e}

\usepackage{graphicx}

\title[
LL-GRBs as a Distinct GRB Population
]
{Low-Luminosity Gamma-Ray Bursts as a Distinct GRB Population:
A Firmer Case from Multiple Criteria Constraints
}
\author[Virgili, Liang, and Zhang]
{Francisco J. Virgili$^{1}$\thanks{virgilif@physics.unlv.edu},
En-Wei Liang$^{2}$ \thanks{lew@physics.unlv.edu},
and Bing Zhang$^{1}$\thanks{bzhang@physics.unlv.edu}\\
$^{1}$Department of Physics and Astronomy,
University of Nevada Las Vegas, Las Vegas, NV 89154, USA\\
$^{2}$Department of Physics, Guangxi University, Nanning 530004, China}
\begin{document}

\date{Submitted . Received 2008 ; in original form 2008}

\pagerange{\pageref{firstpage}--\pageref{lastpage}} \pubyear{2008}

\maketitle

\label{firstpage}

\begin{abstract}
The intriguing observations of {Swift}/BAT X-ray flash XRF 060218
and the BATSE-BeppoSAX gamma-ray burst GRB 980425, both with much
lower luminosity and redshift compared to other observed bursts,
naturally lead to the question of how these low-luminosity (LL)
bursts are related to high-luminosity (HL) bursts. Incorporating
the constraints from both the flux-limited samples observed with
CGRO/BATSE and Swift/BAT and the redshift-known GRB sample, we
investigate the luminosity function for both LL- and HL-GRBs
through simulations.  Our multiple criteria, including  the $\log N - \log P$
distributions from the flux-limited GRB sample, the redshift and
luminosity distributions  of the redshift-known sample, and the
detection ratio of HL- and LL- GRBs with Swift/BAT, provide a set
of stringent constraints to the luminosity function. Assuming that
the GRB rate follows the star formation rate, our simulations show
that a simple power law or a broken power law model of luminosity
function fail to reproduce the observations, and a new component
is required. This component can be modeled with a broken power,
which is characterized by a sharp increase of the burst number at
around $L < 10^{47}$ erg s$^{-1}$. The lack of detection of
moderate-luminosity GRBs at redshift $\sim 0.3$ indicates that
this feature is not due to observational biases. The inferred local rate,
$\rho_0$, of LL-GRBs from our model is $\sim 200$ Gpc$^{-3}$
yr$^{-1}$ at $\sim 10^{47}$ erg s$^{-1}$, much larger than that of
HL-GRBs. These results imply that LL-GRBs could be a separate GRB
population from HL-GRBs. The recent discovery of a local X-ray
transient 080109/SN 2008D would strengthen our conclusion, if the
observed non-thermal emission has a similar origin as the prompt
emission of most GRBs and XRFs.

\end{abstract}


\section{Introduction}
Long duration Gamma-ray bursts (GRBs) are believed to be tied to the
death of massive stars (Colgate 1974; Woosley 1993; Stanek et
al. 2003; Hjorth et al. 2003; Campana et al. 2006).  Of the roughly
6000 bursts observed since the late 1960s (see e.g.
http://heasarc.gsfc.nasa.gov/grbcat/), 
almost 100 have redshift measurements.
Observations show that long GRBs are scattered over a large redshift
and luminosity\footnote{Throughout the text the burst luminosity
refers to the isotropic equivalent value, which does not include the
possible correction of the unknown beaming factor of the GRBs.} 
range, from $z=0.0085\sim$ 6.7\footnote{This record redshift measurement is from GRB 080913.  It remains unclear what category this particular burst belongs to, Type I or Type II, and is not included in the statistical analysis of this work.} and $L=10^{46}\sim
10^{54}~{\rm erg~s}^{-1}$. Most of these bursts have high luminosity
(HL, $L>$ several $10^{48}$ erg s$^{-1}$) with the exception of two
peculiar bursts, GRBs 980425 and 060218, which have extremely low
redshift and luminosity measurements, $(z,L) = (0.0085, 4.7\times
10^{46}~{\rm erg~s}^{-1}$) and (0.033, 6.03$\times 10^{46}~{\rm
erg~s}^{-1}$) respectively (Tinney et al. 1998; Mirabal et al 2006).
It remains unclear whether the LL-GRBs are due to unusual progenitor
properties or a unique population with an intrinsic difference in the
central engine, i.e., black hole versus magnetar (Mazzali et al. 2006;
Soderberg et al. 2006; Toma et al. 2007). 

One empirical way to look into this problem is to see whether
observational data collected so far are still consistent with
LL-GRBs as a natural extention of HL-GRBs to low luminosities in a
continuous luminosity function (LF), or if LL-GRBs form a distinct
new LF component. We have suggested that the latter possibility
(two-component LF model) is necessary based on the redshift-known
sample of GRBs after the discovery of GRB 060218 (Liang et al.
2007). Such a possibility has also been considered as a hypothesis
in the BeppoSAX era (Coward 2005). On the other hand, although
Guetta \& Della Valle (2007) agreed that the two-component model
is possible, they also argued that the $z$-known GRB sample may be
also consistent with a single component model with a steeper slope
in the luminosity function so that more LL-GRBs are accounted for.

Comparing observational data with simulations is a useful way to 
address the relationship between LL-GRBs
and HL-GRBs. It is an important task to constrain the luminosity
function, $\Phi(L)$, and local rate of GRBs, $\rho_0$, in a manner
that can self-consistently reproduce various observations. In
particular, the current LL-GRB population studies have been
focused on the $z$-known sample only. They were not confronted
with the existing $\log N - \log P$ distribution of the BATSE GRB
sample. The $\log N-\log P$ distribution (or $V/V_{\rm max}$
distribution) addresses the statistical properties of GRBs
regardless of their redshift, and carries essential information of
the GRB LF. For HL-GRBs, this criterion has been utilized
extensively (e.g. Schmidt 2001; Stern et al. 2002; Lloyd-Ronning
et al. 2002; Norris 2002; Guetta et al. 2005) and was confronted
with simulations (Lloyd-Ronning et al. 2004; Dai \& Zhang 2005;
Daigne et al. 2006). The conclusion has been that $\Phi(L)$ of
HL-GRBs is generally characterized by a one-component broken power
law model with $\rho_{0, HL}\sim 1 ~{\rm Gpc^{-3}~yr^{-1}}$ (e.g.
Schmidt 2001; Guetta et al. 2004, 2005). On the other hand, the
$\rho_0$ of LL-GRBs inferred from the two detections (GRBs 980425
and 060218) in a decade suggests a much higher local rate than
that of HL-GRBs, i.e., $\rho_{0,LL}=100 \sim 1000$ Gpc$^{-3}$
yr$^{-1}$ (Coward 2005; Cobb et al. 2006; Pian et al. 2006;
Soderberg et al. 2006; Liang et al. 2007; Chapman et al. 2007).
Guetta et al. (2004) propose an extension to lower luminosities
which would increase ${\rho_{\rm 0, HL}}$ from roughly $1.1~{\rm
Gpc^{-3}~yr^{-1}}$ to $10~{\rm Gpc^{-3}~yr^{-1}}$, at roughly
$10^{48}~{\rm erg~s^{-1}}$. This local rate, however, even
extrapolated down to $10^{45}~{\rm erg~s^{-1}}$ is not
sufficiently large to produce the observed LL events. This was the
main motivation of the two-component model in our previous
analysis (Liang et al. 2007, see also Coward 2005; Le \& Dermer
2007). The analysis with the $z$-known sample makes an arguable
case (Liang et al. 2007), but is by no means conclusive (cf.
Guetta \& Della Valle 2007).

In this paper we extend our previous analysis to include a more
complete set of observational constraints. In particular, we
introduce the important BATSE and Swift $\log N-\log P$ distribution
criteria along with the previously considered multiple criteria
involving the $z$-known sample (1-D $z$-distribution, 1-D
$L$-distribution, 2-D $z-L$ distribution, and the observed number
ratio of HL- vs. LL-GRBs). Although the number of LL-GRBs remains
the same, the $z$-known sample has grown since our last analysis
in Liang et al. (2007),  a firmer conclusion drawn in this paper
is largely due to the additional observational criterion included
in this analysis. In order to confront multiple criteria with
different LF models and a wide range of LF parameters, we utilize
a series of Monte Carlo simulations (MCSs). Instrument
observational selection effects are difficult to model, and we
introduce some empirical formulae to roughly reflect gamma-ray
detector trigger sensitivity and the selection effect of redshift
measurement. Various models are presented in \S 2. Our siumlation
results are shown in \S 3, and the conclusions and discussion are
presented in \S 4. The concordance cosmology with parameters
$H_{\rm 0} = 71$ km s$^{-1}$ Mpc$^{-1}$, $\Omega_{\rm{m}}=0.3$
and $\Omega_\Lambda = 0.7$ is assumed throughout.

Throughout the paper we do not touch on another distinctly different 
group of bursts, namely short-hard (Kouveliotou et al. 1993), or 
more general Type I (see e.g. Zhang et al. 2007 for a
discussion of the multiple criteria needed to classify GRBs) bursts,
which are found to be consistent with the compact-star-merger origin
(Gehrels et al. 2005; Fox et al. 2005; Barthelmy et al. 2005;
Berger et al. 2005c, see M\'esz\'aros 2006, Nakar 2007, 
Zhang 2007 for reviews). The analysis of these GRBs applying 
the same technique will be presented elsewhere.

\section{Models}
\subsection{Number of detectable GRBs with an Instrument}
 Assuming that the GRB rate at redshift $z$ is
$R_{GRB}(z)$ (number of GRBs per unit time per unit volume), the
number of GRBs happening per unit (observed) time in a comoving
volume element $dV(z)/dz$ is
\begin{equation}\label{dn}
\frac{dN}{dtdz}=\frac{R_{GRB}(z)}{1+z}\frac{dV(z)}{dz},
\end{equation}
where the $(1+z)$ factor accounts for the cosmological time
dilation, and $dV(z)/dz$ is given by
\begin{equation}\label{volume}
\frac{dV(z)}{dz}=\frac{c}{H_{\rm 0}}\frac{4\pi D_{L}^2}{(1+z)^2
[\Omega_M(1+z)^3+\Omega_\Lambda]^{1/2}},
\end{equation}
for a flat $\Lambda$CDM universe.  The observed GRB/supernovae
connection suggests that the GRB rate could roughly trace the star
formation history\footnote{More recently, some authors suggested
that the observed GRB rate at high redshift is higher than the
star formation rate (Kistler et al. 2007; Daigne et al. 2007; Li
2007; Cen \& Fang 2007). We will explore various redshift-dependent
effects in a future work.}.
We adopt a parameterized GRB rate model proposed by
Porciani and Madau (2000),
\begin{equation}\label{SF2}
R_{GRB}=23\rho_{\rm0}\frac{e^{3.4z}}{e^{3.4z}+22.0}.
\end{equation}
or by Rowan-Robinson (1999),
\begin{equation}
\label{RR}
R_{GRB}= \rho_0 \left\{
\begin{array}{l@{\quad \quad}l}
10^{0.75z} & z<1\\
10^{0.75z_{peak}} & z>1,\\
\end{array}
\right.
\end{equation}
where $z_{peak}$ is the redshift at which the redshift
distribution reaches its maximum (after which it plateaus),
taken here as 1.

Supposing the GRB luminosity function is $\Phi(L)$, the number of GRBs per unit
time at redshift $z\sim z+dz$ and luminosity $L\sim L+dL$ is given by
\begin{equation}\label{dN}
\frac{dN}{dtdzdL}=\frac{R_{GRB}(z)}{1+z}\frac{dV(z)}{dz}\Phi(L).
\end{equation}

Considering an instrument with energy band [$e_{1}$, $e_2$] having
a flux threshold $F_{\rm th}$ and an average solid angle $\Omega$
for the aperture flux, the number of the detected GRBs during an
observational period of $T$ should be
\begin{equation}\label{N}
N=\frac{\Omega T}{4\pi}\int_{L_{1}}^{L_{2}}
\Phi(L)dL\int_0^{z_{\max}} \frac{R_{\rm
GRB}(z)}{1+z}\frac{dV(z)}{dz}dz,
\end{equation}
where $z_{\max}$ for a given burst with luminosity $L$ is
determined by the instrumental flux threshold $F_{\rm th}$ through
$F_{\rm th}=L/4\pi D^2_L(z_{\max})k$. The $k$ factor corrects the
observed flux in an instrument band to bolometric flux in the
burst rest frame ($1-10^4$ keV in this analysis),
\begin{equation}\label{kcorr}
k=\frac{\int_{1/(1+z)}^{10^4/(1+z)}EN(E)dE}{\int_{e_1}^{e_2}EN(E)dE}.
\end{equation}
where $N(E)$ is the photon spectrum of GRBs. It is generally
fitted with a joined power law (the Band function; Band et al.
1993) characterized with photon indices $\Gamma_1$ and $\Gamma_2$
before and after a break at $E_0$. The peak energy of the $\nu
f_\nu$ spectrum is given by $E_p=E_0(2+\Gamma_1)$. It was shown
that $\Gamma_1\sim -1$, $\Gamma_2\sim -2.3$, and $E_p\sim 250$ keV
for a typical GRB (Preece et al. 2000). In our analysis, the
luminosity extends over ten orders of magnitude ([$10^{45}$,
$10^{55}$] erg $\rm s^{-1}$). According to the Amati relation
(Amati et al. 2002; Liang et al. 2004), more luminous bursts have
a higher $E_p$, indicating that we cannot adopt a uniform $E_p$
for all the bursts in our analysis. Liang et al. (2004) derived
\begin{equation}\label{EpLiso}
 E_p/200 {\rm keV}=C (L/10^{52} {\rm erg\ s}^{-1})^{1/2}
\end{equation}
where $C$ is randomly distributed in [0.1,1]. We obtain $E_p$ with
Eq. \ref{EpLiso} for each burst and assume $\Gamma_1\sim -1$ and
$\Gamma_2\sim -2.3$ for all bursts.

With the spectral information, one can make the k-correction and
get the observed peak energy flux and peak photon flux by
\begin{equation}\label{Flux}
F=\frac{L}{4\pi D_L^2 k}
\end{equation}
and
\begin{equation}\label{Photon}
P_{ph}=\frac{F\int_{e_1}^{e_2}N(E)dE}{\int_{e_1}^{e_2}EN(E)dE},
\end{equation}
respectively.

\subsection{Luminosity Functions}
Attempts to determine $\Phi(L)$ of long GRBs  have been made by
some authors, through fitting the $\log N-\log P$ or $V/V_{\rm
max}$ distributions observed by CGRO/BATSE (Schmidt 2001; Stern et
al. 2002; Lloyd-Ronning et al. 2002; Norris 2002; Guetta et al.
2005), and generally characterize $\Phi(L)$ with a single
power-law or an a broken power law within a given luminosity range
[$L_1$, $L_2$], i.e.,
\begin{equation}\label{G04}
\Phi(L)=\Phi_0\left(\frac{L}{L_B}\right)^{-\alpha},
\end{equation}
or
\begin{equation}\label{LF}
\Phi(L)=\Phi_0\left[\left(\frac{L}{L_b}\right)^{\alpha_1}
+\left(\frac{L}{L_b}\right)^{\alpha_2}\right]^{-1},
\end{equation}
where $\Phi_0$ is a normalization constant to assure
$\int_{L_1}^{L_2}\Phi(L) dL=1$. The local GRB rate, $\rho_0
=R_{GRB}(z=0)$, is in principle defined to include GRBs with all
luminosities. In practice, since observations cannot probe the
full luminosity function, the $\rho_0$ value constrained by the
data is usually related to a lowest luminosity $L_1$. The value of
$\rho_0$, therefore, is a function of $L_1$. For a single power
law LF (Eq. \ref{G04}) with $\alpha > 1$, one has $\rho_0(L>L_1)
\propto L_1^{-(\alpha-1)}$, suggesting that a lower $L_1$ would
give rise to a larger observed $\rho_0(L>L_1)$. For a broken power
law LF (Eq.\ref{LF}) with $\alpha_1 < 1$ and $\alpha_2 > 1$, on
the other hand, integration suggests that $\rho_0(L>L_1) \sim
\rho_0(L>L_b)$ which is essentially independent of $L_1$. Thus
fixing $L_b$ would usually fix $\rho_0$ in the broken PL models.
In the past, the LF of HL-GRBs was found to have a break around
$L_b \sim 10^{50}\rm~erg~s^{-1}$, with the value of $\rho_0$
related to $L_b$. In our analysis, the local rate is evaluated at
a lower luminosity cutoff for all models, although the value is
determined by either $L_1$ or $L_{\rm b,LL}$ depending on the
forms of LF adopted. These are summarized in Table 1.

\subsection{Instrument Threshold and Detection Biases}
In order to check if a simulated burst is detectable with a given
instrument, the simulated burst is screened with the instrument
threshold. The CGRO/BATSE was triggered by energy-dependent count
rate (Band 2003). We take a moderate sensitivity for CGRO/BATSE at
50-100 keV band as $F_{th}^{BATSE}\sim 10^{-7}$ erg cm $^{-2}$
s$^{-1}$ , roughly corresponding to $0.2$ ph cm$^{-2}$ s$^{-1}$
for a typical GRB.

The BAT instrument on board {\it Swift} operates with an image
trigger mechanism. The sensitivity of an event depends on many
complicated factors and in principle should be treated in the
case-by-case basis. For the purpose of this paper, we adopt an
approximate formula of Sakamoto et al. (2007),
\begin{equation}\label{Fth}
F_{th} \sim {(5.3\times10^{-9}~\rm erg~cm^{-2}~s^{-1})} {f}^{-1}
t_{90}^{-0.5}.
\end{equation}
where $f$ is the partial coded fraction and $t_{90}$ is the
burst duration.
The larger the burst duration, the more sensitive the instrument.
Bursts with $F>F_{th}$ can be in principle detected. Observationally,
the ``peak fluxes'' (and therefore the ``peak luminosities'') are
usually adopted to denote for the brightness of the bursts, which
are on average 5 times above the average fluxes 
(T. Sakamoto, private communication). To compensate this
effect, we hereafter adopt an effective threshold condition which
is 5 times larger than Eq.(\ref{Fth}) in our peak-luminosity
analyses. Since HL-GRBs have a typical duration of 20 s, we adopt
a rough constant threshold flux of $F_{th,eff}\sim 1.2 \times10^{-8}~\rm
erg~cm^{-2}~s^{-1}$ for the analyses of HL-GRBs. As shown by Norris
et al. (2005), LL-GRBs tend to have longer pulse duration. GRB 060218,
for example, has a duration longer than 2000 seconds (Campana et
al. 2006; Liang et al. 2006). In our analysis we adopt various discrete
values of BAT sensitivities, not exceeding a
500 second duration (i.e., $4.7\times10^{-10}~\rm
erg~cm^{-2}~s^{-1}$), to screen the simulated LL-GRBs.

Theoretically, a GRB could be detectable if $F>F_{th,eff}$. Note that
a large fraction of detectable events with $F$ close to $F_{th,eff}$
may not trigger the instrument. This fact was observed in
CGRO/BATSE. An off-line scan found a large number of non-triggered
GRBs in the BATSE catalog, most of them are near the instrument
threshold (Stern et al. 2001). Our simulations, by not adopting a
threshold for the $\log N-\log P$ distribution have an advantage in
deciphering the intrinsic low photon flux end of the distribution,
which may be tested in the future by more seneitive detectors such as
JANUS (Roming et al. 2008) and EXIST (Grindlay 2006).

In our analysis we also compare the simulated sample with the
redshift-known GRB sample. This sample suffers many observational
biases (Bloom et al. 2001; Butler et al. 2007), including
position localization, optical detection, and line detection.
The trigger probability of a burst with a flux
close to the instrument threshold tends to be low, as seen in
BATSE. Near-threshold GRBs tend to have fainter optical
afterglows, which severely bias against their redshift
measurements. It is difficult to fully incorporate all these
biases into the simulation. We simply model the redshift detection
probability of a simulated burst based on an empirical formula,
\begin{equation}
p(F)=(1-\frac{F_{th}}{F})^\kappa, \label{trigger prob}
\end{equation}
where $\kappa$ is a free parameter. Our analyses show that $\kappa
\sim 7$ is necessary to eliminate the overproduction of bursts
near the sensitivity threshold.  Equation \ref{trigger prob} is not 
highly sensitive on the value of $\kappa$, whose most notable effect 
is on bursts near the threshold.  As the value of $\kappa$ is 
decreased, most bursts appear near the threshold and below the band of observed bursts, significantly decreasing the significance of the correlation between the simulated and observed bursts. We notice that the correlation between gamma-ray and optical luminosities is not a strong one (e.g. Liang \& Zhang 2006; Nardini et al. 2006; Kann et al. 2007; Nysewander et al. 2008), and that determination of redshift is easier in some ranges than others. All these make the selection effects more complicated than the simple parameterization such as Eq.\ref{trigger prob}. Nonetheless, without simulating the optical luminosities, here we take a simple form for the sake of simplicity, which can effectively screen low-luminosity bursts without affecting the bursts significantly above the threshold.

\section{Monte Carlo Simulations Against Observations}
We place constraints on the parameters of $\Phi(L)$ and $R_{GRB}$
through comparing our simulations with observations. The primary
criterion to judge the parameters is that the simulated $\log N -
\log P$ distribution should match the data. The detection number
of an instrument in a given observation period should roughly
match the observations. CGRO/BATSE and Swift/BAT established two
uniform samples that can constrain the parameters with the
detection event number and the fits to the observed $\log N-\log
P$ distributions. The size of the mock GRB sample for a given
instrument accumulated in a period $T$ is obtained with Eq.
\ref{N}. CGRO/BATSE recorded 1637 triggered long GRB events during 9.1
operation years (4B catalog, Paciesas, et al., 1999), and Swift/BAT
was triggered by $\sim 300$ GRBs during the first 3 years of
operation.
\footnote{We don't include those non-triggered
events (Schmidt 2004).}
We regard all detectable type II GRB events that could
trigger BAT. Therefore, a set of the parameters can be obtained by
adjusting the parameters that makes the detectable event numbers
with BATSE and BAT could be $2176$ events in $9.1$ years (1393 triggered and 874 non-trggered) and
$N=300$ in $3$ years, respectively, and the observed $\log N-\log
P$ with the two instruments that match our simulations.  We measure the
consistency between observations and our simulations by a K-S test,
resulting in a probability $p_{\rm K-S}$ (Press et al. 1997).
The larger $p_{\rm K-S}$ suggests a more significant consistency.

The second criterion to judge the parameters is the constraints
from the redshift-known sample and the detections of LL-GRBs. The
current GRB sample with redshift measurements has $\sim 100$ GRBs,
and only two confirmed detections of LL-GRBs. They were detected by
different instruments. Although the sample is not homogeneous for
statistics, the parameters of $\Phi(L)$ and $R_{GRB}$ are subject
to the constraints from this sample, especially when we take the
LL-GRBs into account (Liang et al. 2007). First, the detection
ratio of LL-GRBs to HL-GRBs should be $\sim 1:300$ as observed by
BAT in three years. Second, both the simulated one-dimensional
distributions of $L$ and $z$  and the GRB distribution in the
$(L-z)$ plane should roughly match the observations. We therefore
simulate a sub-sample of GRBs that have redshift measurement based
on the probability of redshift measurement (Eq. \ref{trigger
prob}) from simulated GRB sample for BAT. We measure the
one-dimensional $L$ and $z$ distribution with the K-S test, and
then combine the distributions in order to place more
rigorous constraints. We compare two-dimensional contributions in
the $L-z$ plane and measure the consistency of the simulated
sample to the observational data with $p_{\rm K-S,t}=p_{\rm K-S,L}
\times p_{\rm K-S,z}$.

With the criteria described above, we adjust the model parameters
and simulate a large sample of GRBs and filter them with a
given instrument threshold without considering the cosmological
evolution of $\Phi(L)$, stopping the simulations when a sub-sample
of 150 detectable bursts are reached.  Our simulation procedure is
summarized as follows.

First, we simulate a burst that is characterized with $L$ at
redshift $z$, GRB($L,z$). Both quantities are simulated separately
with the probability distributions derived from Eq. \ref{dn} and
$\Phi(L)$ (one of Eqs. \ref{G04}-\ref{LF}).

Second, we calculate the $E_p$ of a mock GRB with Eq. \ref{EpLiso}, and its $F$ and $P_{ph}$ from Eqs. \ref{Flux} and \ref{Photon},
respectively. The simulated GRB then is screened with the
threshold condition $F>F_{th}$.

Third, we simulate a redshift-known sub-sample from the
flux-cutoff mock sample with Eq. \ref{trigger prob}.

In each of the following subsections we will try different LF
models and compare the simulations with the observations.

\subsection{Simple Power Law Model}
The first model considered is the simple power law model (Eq. \ref{G04}). This
scenario has been extensively studied with the BATSE data (see \S
1). Guetta et al. (2004) investigate this model without
considering the LL-GRBs. In order to explain the high detection
rate of LL-GRBs, Guetta \& Della Valle (2007) proposed that the
GRB luminosity function is a single PL with a slope $\alpha=1.6$
and $\rho_0=1.1$, 200, or (200-1800) Gpc$^{-3}$ yr$^{-1}$
depending on which lower luminosity cutoff was used, $5\times
10^{49}~\rm erg~s^{-1}$, $5\times 10^{47}~\rm erg~s^{-1}$, or
$5\times 10^{46}~\rm erg~s^{-1}$, as summarized in Table 1. We
make simulations with this model and adopt the parameters from
Guetta et al. (2004, 2007).
The simulated results of these models are shown in Figs. 1-4. We
first compare the simulated results to the $\log N- \log P$ distribution observed by BATSE.  In
general, as far as HL-GRBs are concerned, the model is able
to reproduce a good comparison to the $\log N-\log P$ (Fig.1 a,b).
However, in order to accommodate LL-GRBs, modifications of the LF
parameters are needed. The revised model (Guetta \& Della Valle
2007), although able to account for the event rate of LL-GRBs,
deviates from the observed BATSE $\log N - \log P$ distribution
significantly (Fig.1c).
\begin{figure}
\includegraphics[angle=0,scale=0.6]{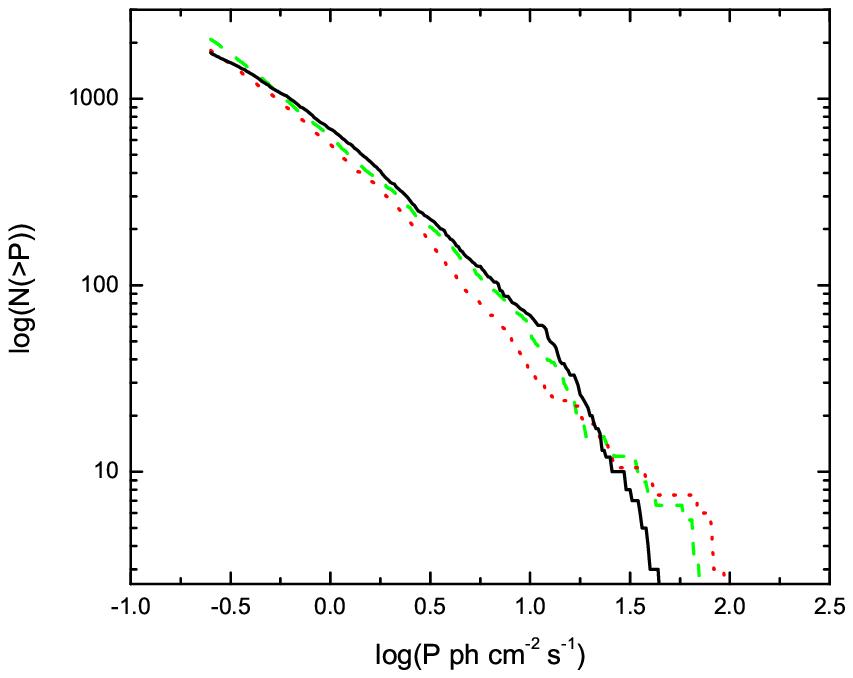}
\includegraphics[angle=0,scale=0.6]{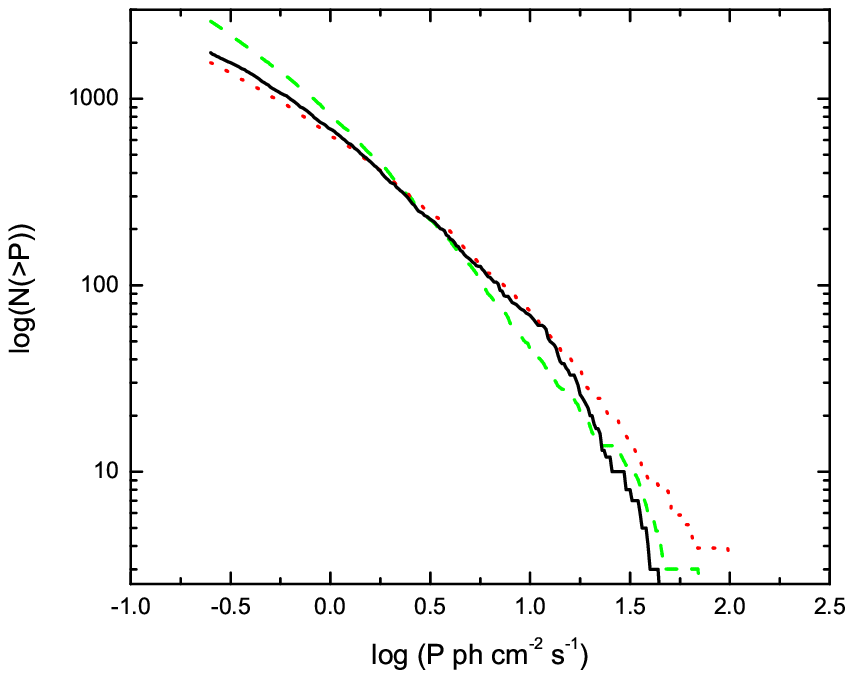}
\includegraphics[angle=0,scale=0.6]{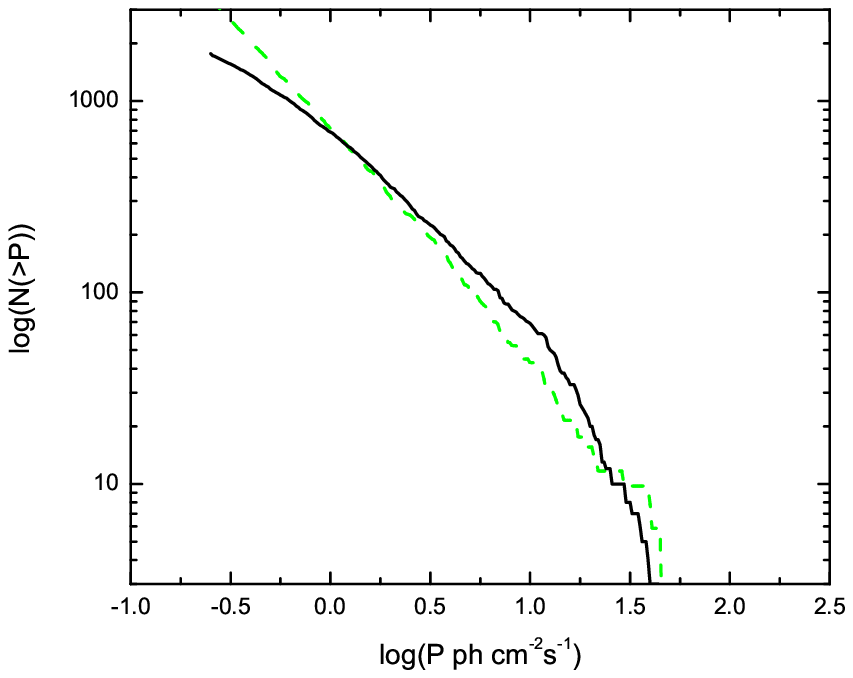}
\caption{One-LF model fits to BATSE $\log N-\log P$ distribution. The
solid line (black) denotes the observed BATSE $\log N-\log P$
distribution in each pane. From left to right (a-c), we have the models
from G04 (G04 (green, dash); G04(2),(red, dotted)) G05 (P\&M (green,dash); RR(red, dotted), and G07 with the largest $L_{min}$.  The first two models (G04, G05) can roughly
reproduce the observation, while the last model (G07) is ruled out by
the data. The observed BATSE distribution is the solid black curve in all panes.}
\label{lnlpg}
\end{figure}

The simulated distributions of $L$, $z$, and $L-z$ in a two
dimensional plane are shown in Fig. 2-4 along with the
observational results. Without considering the LL-GRBs, the models
of Guetta et al. (2004) can roughly produce the observed
1-dimensional $L$ and $z$ distributions. The model of Guetta \&
Della Valle (2007), however, causes a severe overproduction  in
bursts of luminosities $\sim 10^{48}-10^{50} \rm erg~s^{-1}$ at
low redshifts and fails to reproduce bursts with z$>\sim$3 for the
largest lower-luminosity cutoff of $5\times 10^{49} \rm
erg~s^{-1}$ (see also Liang et al. 2007).  The two-dimensional
analysis, shown in Fig. 4, demonstrates this overproduction and
makes note of the deficiency of bursts above $\sim 10^{52}~\rm
erg~s^{-1}$ in all models. When LL-GRBs are considered, the models
of Guetta et al. (2004) are insufficient since they predict a
$\rho_0$ that is too low to account for the observed LL-GRBs.
While the modified model by Guetta \& Della Valle (2007) can
accommodate a sufficiently low low-luminosity cutoff of
$0.1L_{980425}$, the steep slope and low cutoff cause a large
deviation from the observed $\log N-\log P$ distribution as
discussed above.

\subsection{Broken Power Law Model}
A single power law LF model encounters great difficulty in
simultaneously reproducing the observed HL/LL-GRB populations and
the BATSE $\log N - \log P$ distributions. Therefore, we try the
broken power law LF model (\ref{LF}), also
discussed by Guetta et al. (2004,2005).  The results of which
are also shown in Figs. 1-4. The BPL model of G04 provides a
distribution that peaks at around  $z\sim$ 1, matching
observations, as well as producing a similar number of bursts all
around.  As the parameters of the BPL are shifted for G05, the
peak remains similar although the distribution becomes narrower.
Apparently, these results are similar to that of the simple power
law model and cannot explain both HL- and LL- GRBs and the $\log N
- \log P$ distributions.

\begin{figure}
\includegraphics[angle=0,scale=0.6]{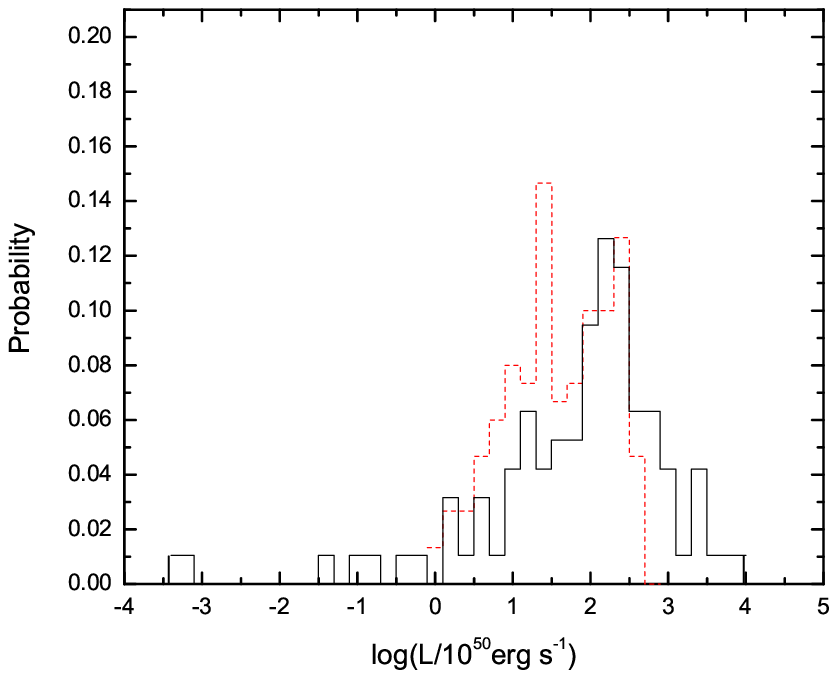}
\includegraphics[angle=0,scale=0.6]{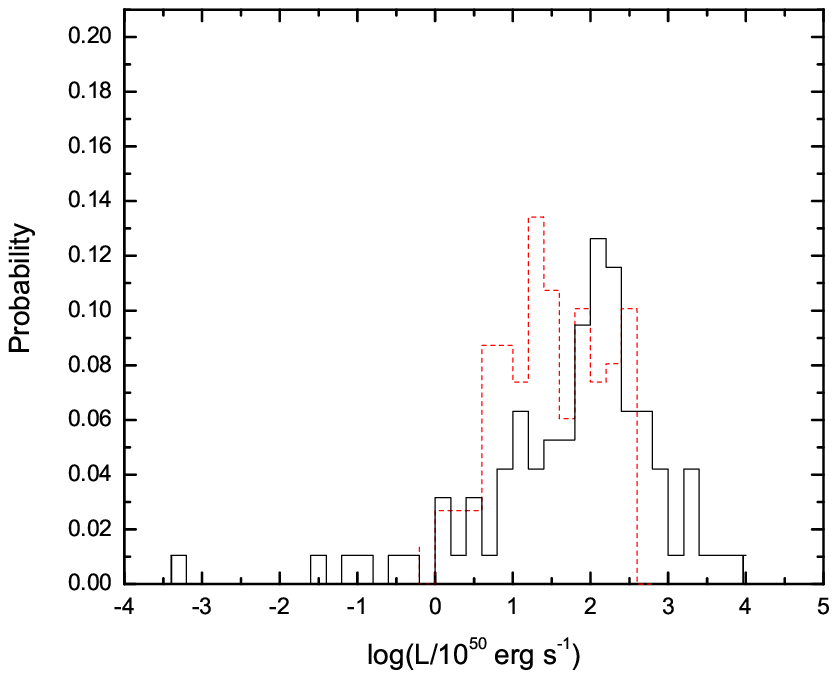}
\includegraphics[angle=0,scale=0.6]{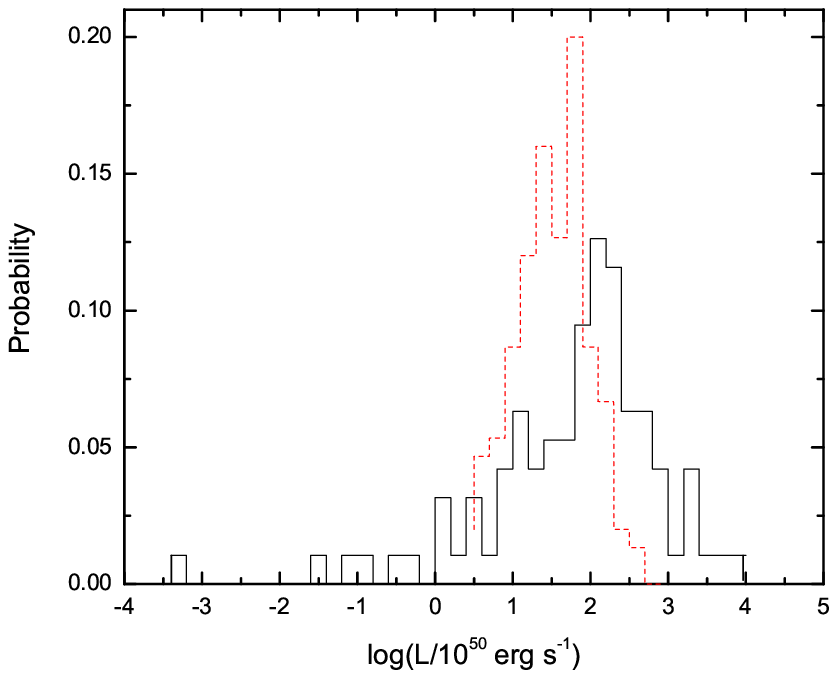}
\includegraphics[angle=0,scale=0.6]{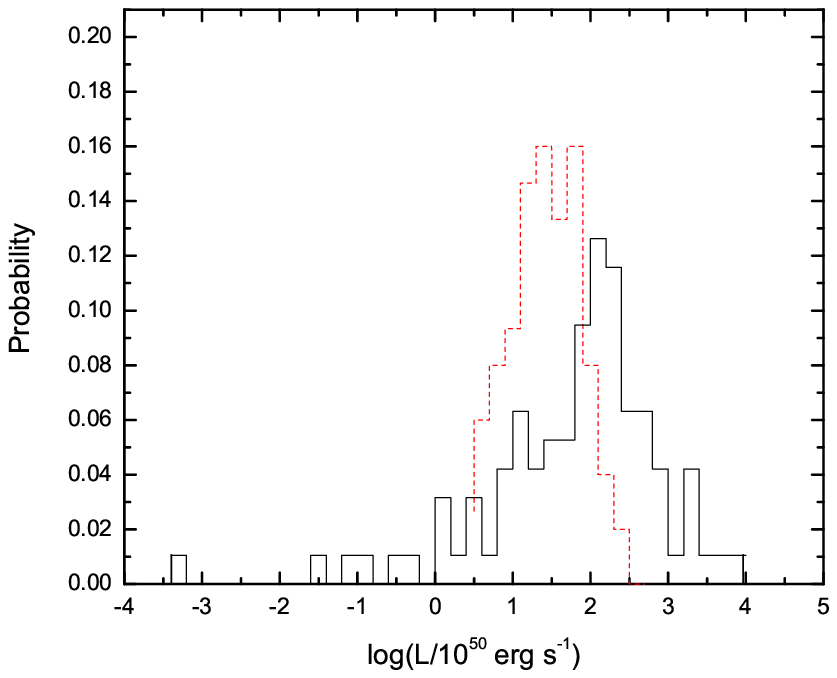}
\includegraphics[angle=0,scale=0.6]{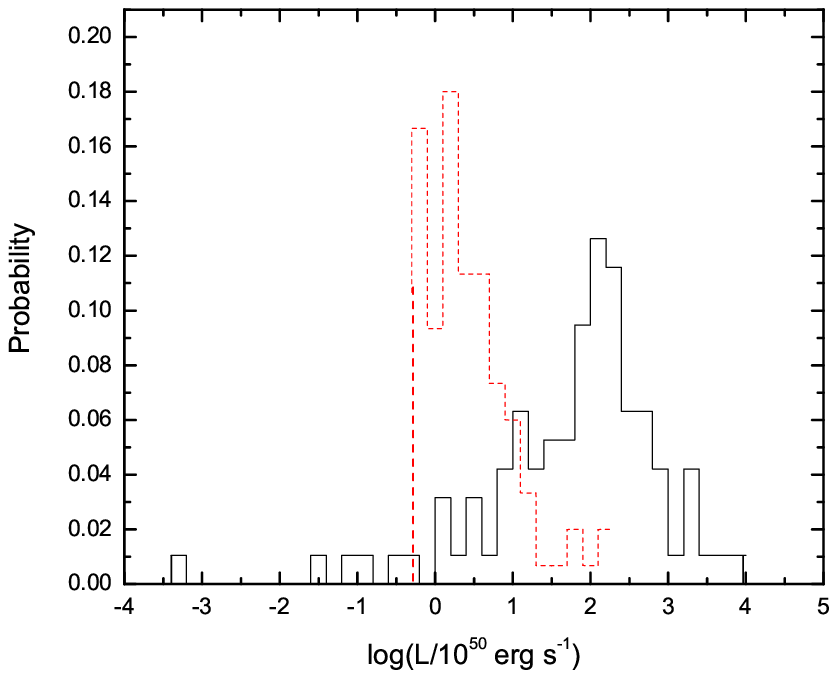}
\includegraphics[angle=0,scale=0.6]{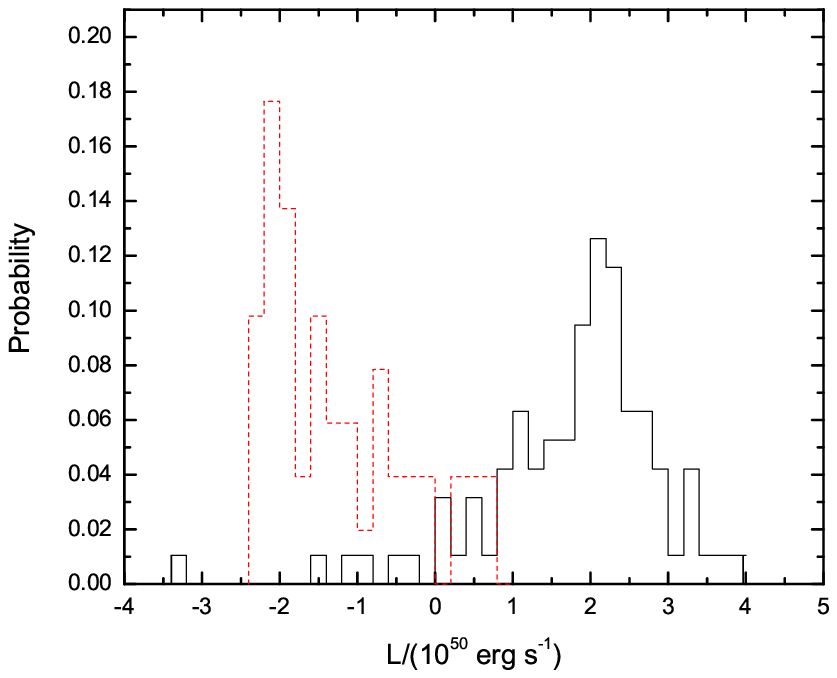}

\caption{The 1-D luminosity distributions of various single-component LF models.
The dashed curves (red) are the simulated results, while the solid curves (black) are
the observed results for the redshift-known sample.  Model parameters can be found
in Table 1. The LF forms are, from left to right, G04, G04(2), G05 (P\&M),
G05 (RR), G07, G07(2).}
\label{Lg}
\end{figure}

\begin{figure}
\includegraphics[angle=0,scale=0.6]{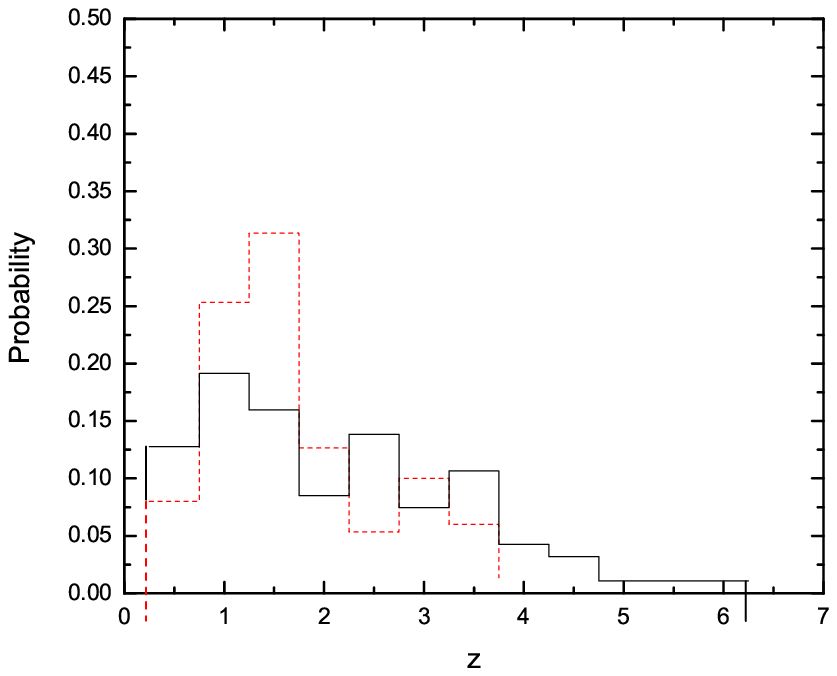}
\includegraphics[angle=0,scale=0.6]{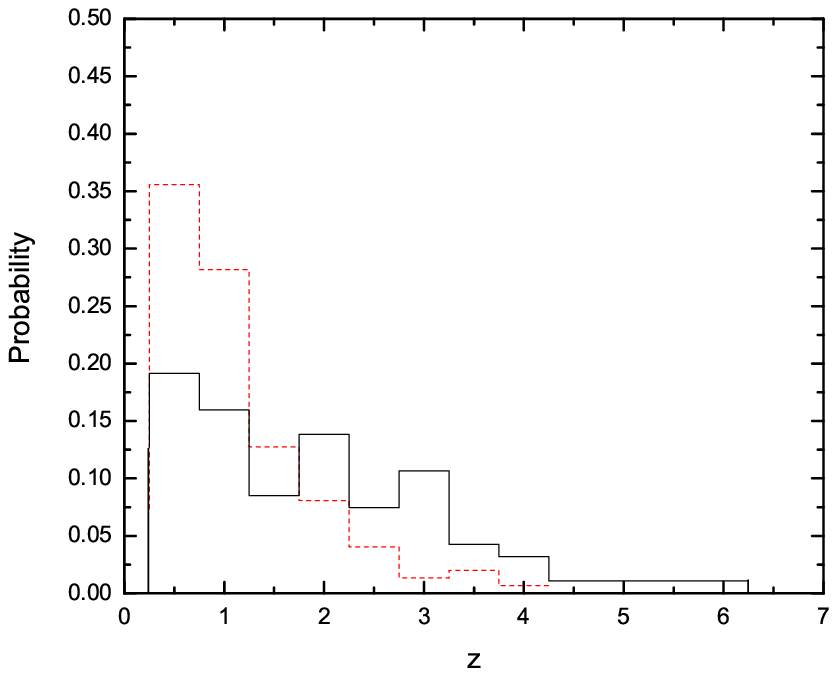}
\includegraphics[angle=0,scale=0.6]{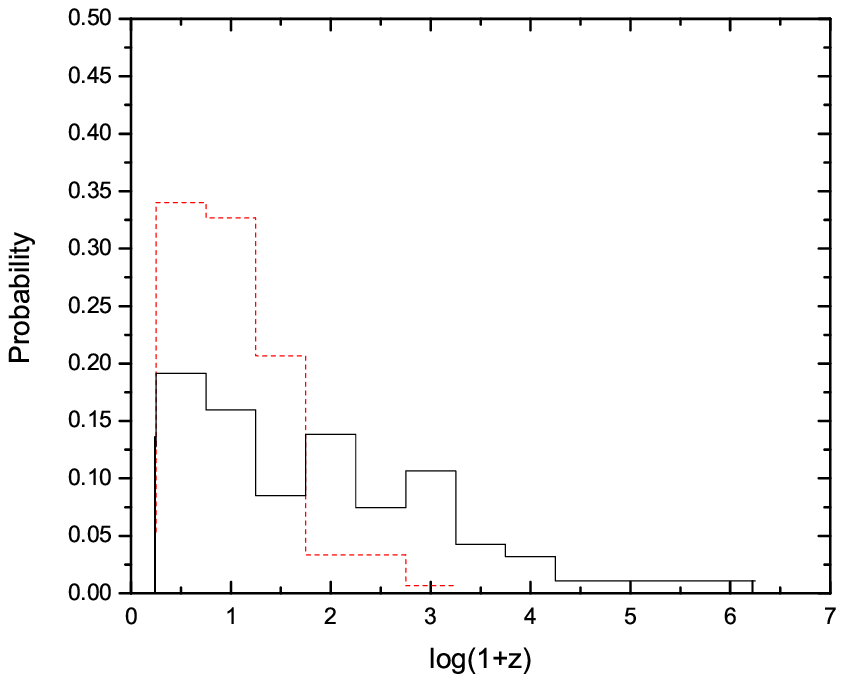}
\includegraphics[angle=0,scale=0.6]{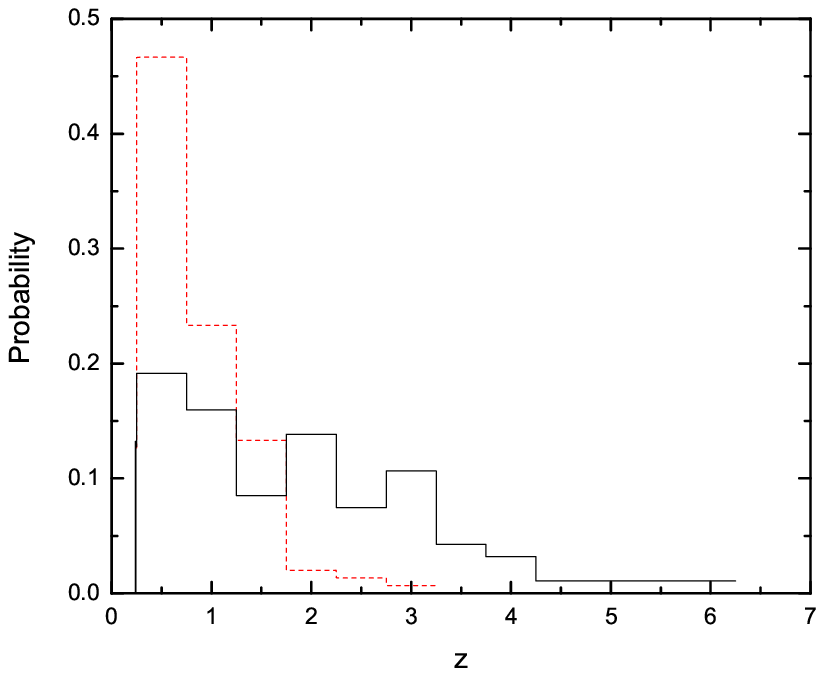}
\includegraphics[angle=0,scale=0.6]{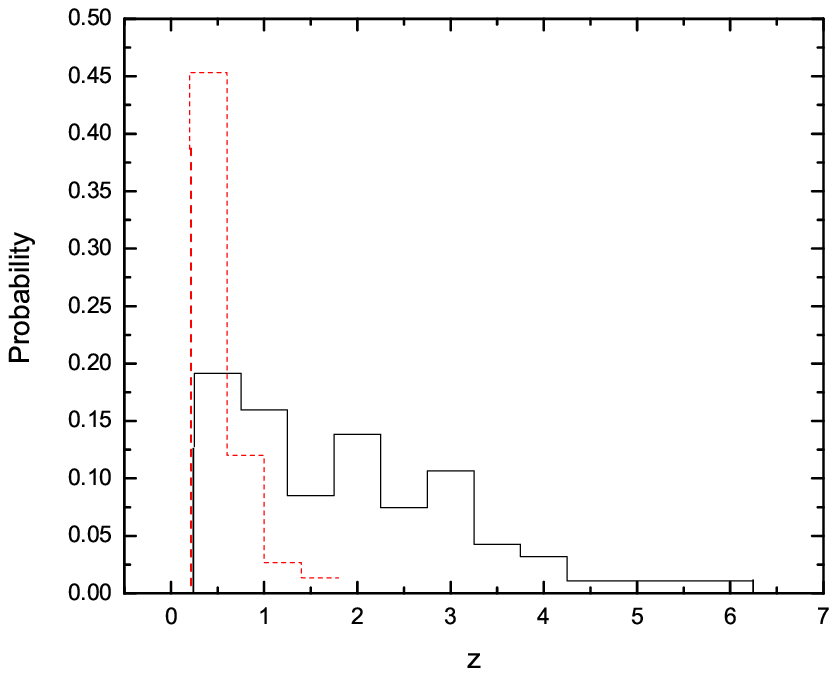}
\includegraphics[angle=0,scale=0.6]{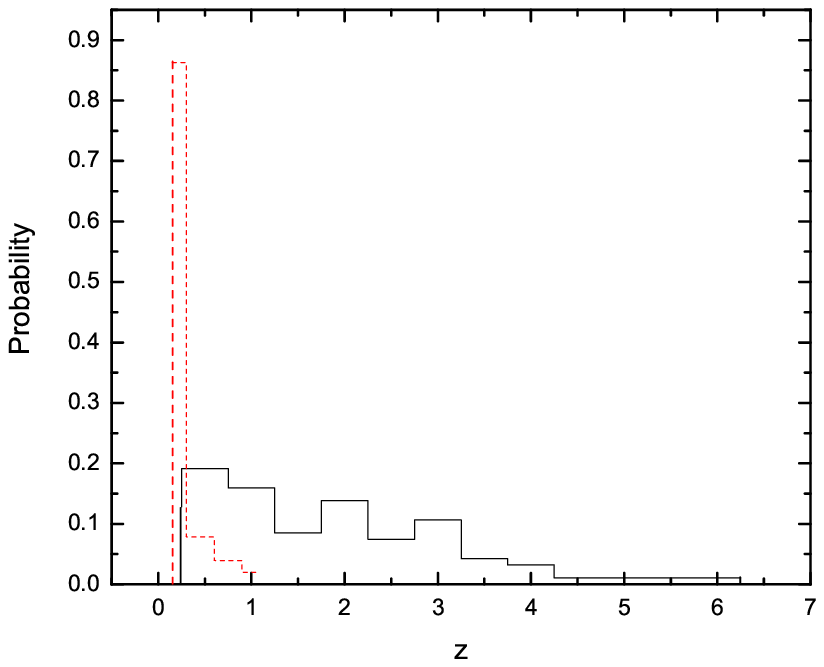}
\caption{The 1-D redshift distributions of various single-component LF models.
The dashed curves (red) indicate simulated results, while the solid curves (black)
indicate the observed results for the redshift-known sample.  The LF forms are,
from left to right, G04, G04(2), G05 (P\&M), G05 (RR), G07, G07(2).}
\label{zg}
\end{figure}

\begin{figure}
\includegraphics[angle=0,scale=0.6]{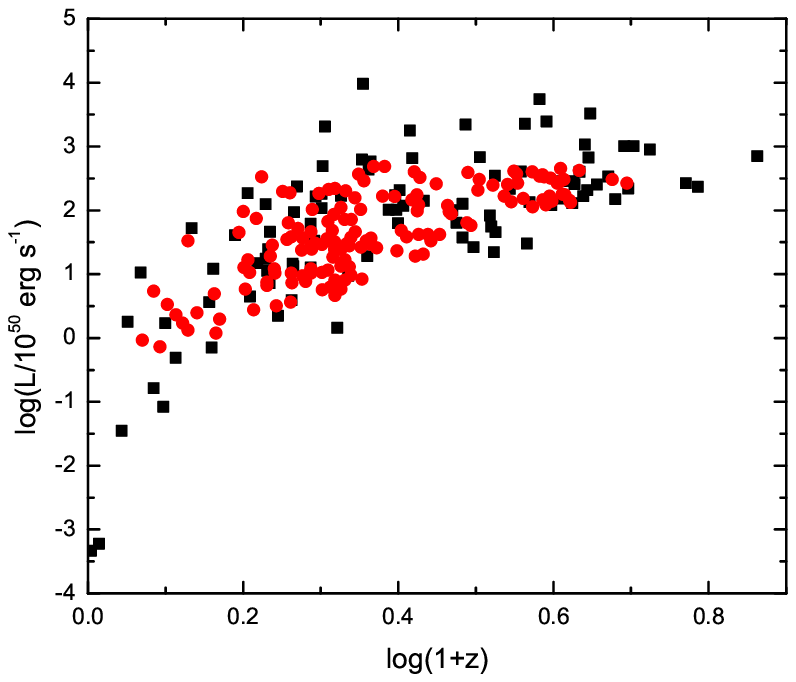}
\includegraphics[angle=0,scale=0.6]{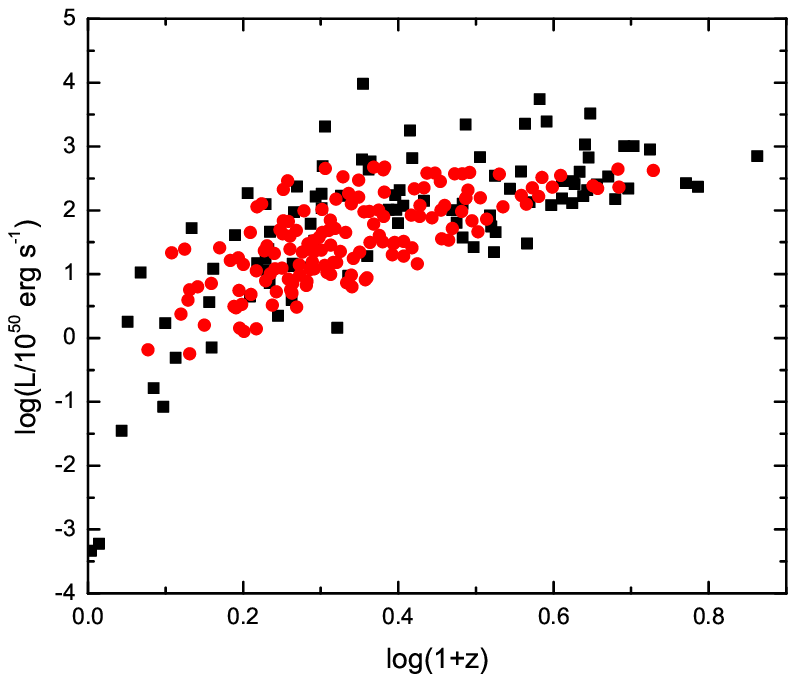}
\includegraphics[angle=0,scale=0.6]{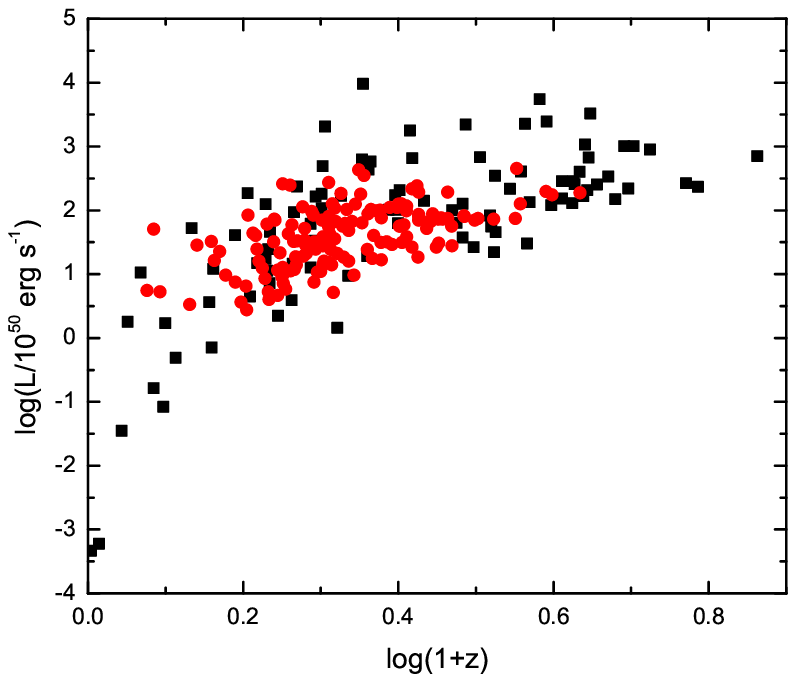}
\includegraphics[angle=0,scale=0.6]{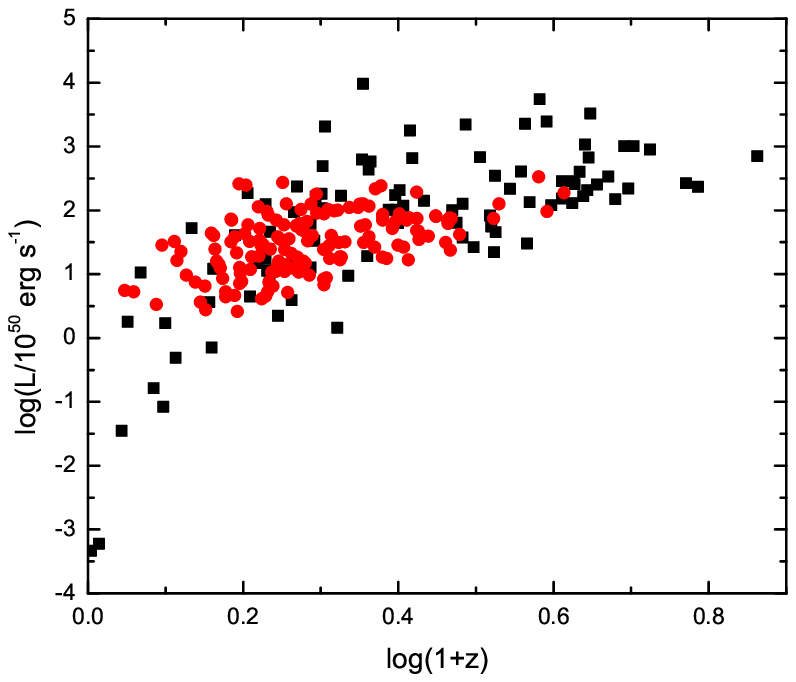}
\includegraphics[angle=0,scale=0.6]{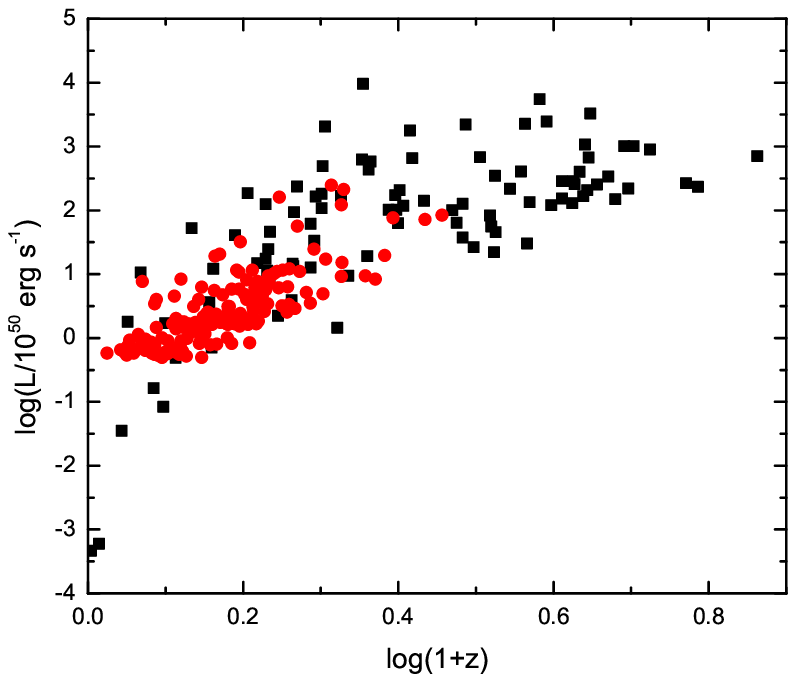}
\includegraphics[angle=0,scale=0.6]{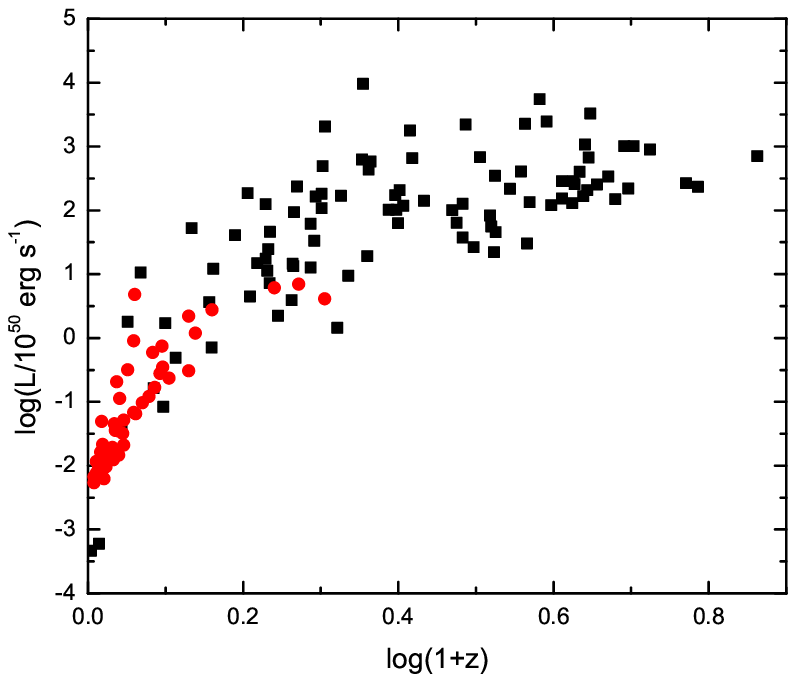}
\caption{Two-dimensional luminosity-redshift distributions of various
single-component PL models.  The filled squares (black) are the observed
redshift-known sample in the $z-L$ plane, while the
filled circles (red) are the simulation results for models various
models.  The LF forms are, from left to right, G04, G04(2), G05 (P\&M),
G05 (RR), G07, and G07(2).  None of these models are able to reproduce
the observed distribution satisfactorily.}
\label{2Dg}
\end{figure}

\subsection{Combined Broken Power law Model: LL-GRBs as a Distinct
GRB Population}

Coward (2005) and Liang et al. (2007) proposed that LL-GRBs could be from a unique
GRB population, characterized by low luminosity, less collimation,
and high local rate compared to HL-GRBs.
With the sensitivity threshold of BATSE and Swift BAT,
these events are only detectable in a small volume, so that the
number of detectable LL-GRBs could be low. With more sensitive
detectors (e.g. JANUS, Roming et al. 2008; EXIST, Grindlay et al.
2006), one can probe into a larger volume which result in a larger
number of detected LL-GRBs.
To date only 2 LL-GRBs (GRBs 980425 and 060218) have been well-localized.
The small number of LL-GRBs
makes statistical testing of this population alone inaccurate.
However, the high local LL-GRB rate inferred from the detection of
GRBs 980425 and 060218 and the deficit of the observed GRBs with
median luminosity ($10^{48}\sim 10^{49} \rm ~erg~s^{-1}$) at
redshift $0.1\sim 0.5$ place strong constraints on the luminosity
function of this GRB population. It is unlikely that the lack of
intermediate-redshift GRBs is the consequence of some selection
effects. Since we are analyzing the intrinsic luminosities (rather
than the observed fluxes), the deficit should not be related to
an instrumental threshold effect, which requires that the LL-GRBs
would diminish along with the intermediate luminosities GRBs.
A redshift-dependent selection effect would require that there
is a strong correlation between burst luminosity and redshift,
which is not discovered from the data. We therefore conclude
that the observations imply an intrinsic feature in the LF.
Liang et al. (2007) suggested that the
global GRB LF can be modeled with two components: a smoothed 
broken power law for each population.  We elaborate this 
two-component model with our simulations.

\begin{figure}
\includegraphics[angle=0,scale=1.5]{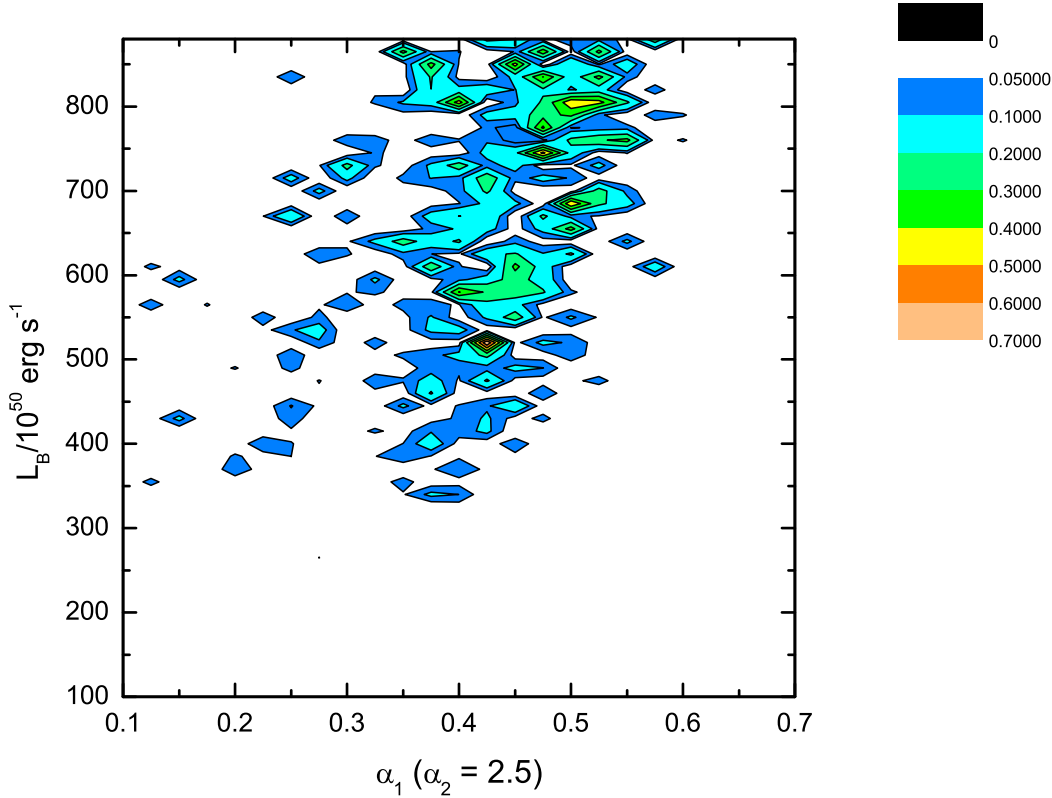}
\caption{2-D $p_{KS,t}$ contour as a function of $\alpha_{\rm 1,HL}$ and $L_b$ at
$\alpha_2 = 2.5$.}
\label{contour}
\end{figure}

The simulated GRBs cover a luminosity interval of
$L=10^{45}-10^{55}\rm~erg~s^{-1}$ and a redshift interval of
$z=0-10$.  The values of luminosity and redshift assigned are
subjected to the detector conditions described above, until a
subset of 300 bursts (similar to the observed number of these
types of bursts by Swift) is achieved. We then constrain the
two-component LF parameters with the following procedure. First,
we vary the parameters for the HL-GRB LF and compare with the
observed 1-D $L-$ and $z-$ distributions as well as the 2-D
$(L-z)$ distributions. Since the results are not sensitive to the
value of $\alpha_{\rm 2,HL}$, we fix it at 2.5, and search for
high-likelihood parameters in the $L_b-\alpha_{\rm 1,HL}$ space
using K-S probability contours (Fig.\ref{contour}). This leads to
high probability concentrations in a variety of spots. We then use
the $\log N-\log P$ criterion to pin down the best parameter
space.

Next, we constrain the LL-GRB component
parameters using the 1-D and 2-D distributions as well as the relative ratio
of the observed HL- and LL-GRBs.
In order to address the number of simulated bursts that pass the threshold
conditions, it is necessary to understand how each population was controlled.
The number of LL bursts created was directly proportional to the number of HL-
bursts created by the ratio of the local rates for each type of burst.  The
number of LL iterations was prescribed by
\begin{equation}\label{number}
N_{LL}=N_{HL}\frac{\rho_{\rm 0,LL}}{\rho_{\rm 0,HL}}.
\end{equation}
A change in either rate will result in different amounts of each
type of burst created, which then affects the final
distributions and observable number of bursts.  It is necessary to note that the observable ratio of about 300:1 (HL:LL) bursts is for all triggered bursts, not just the redshift known subset.  Therefore, the redshift-measurement probability condition (Eq. \ref{trigger prob}) need not be applied to the bursts since the purpose of this addition was to simulate the redshift measurement bias.

Analysis of the two component LF model shows promising results and
constraints on the LF and local rate of both HL- and LL-bursts.
It is found that the parameter set  $(\alpha_{\rm 1,HL},
\alpha_{\rm 2,HL}, L_b)=(0.425, 2.5, 5.2\times 10^{52} \rm
~erg~s^{-1}$) gives a reasonable fit to both the $L-z$ constraints
and the $\log N - \log P$ distribution, producing a strong peak in
$L-z$ probability and fit to the $\log N-\log P$ distribution.
For completeness, we also consider two other sets of parameters,
$(\alpha_{\rm 1,HL}, \alpha_{\rm 2,HL}, L_b)=(0.5, 2.5, 8\times
10^{52} \rm ~erg~s^{-1}$) and ($0.475, 2.5, 7.2\times 10^{52} \rm
~erg~s^{-1}$), which correspond to the second highest peak in
$\alpha_1-L_B$ space, and a sample near the center of the maximum
values of the contour.  These parameter sets are the combination
and result of the first three criteria for constraining the LF
parameters, as mentioned in the introduction.

Moving on to the last criterion, the best fit parameters give number ratios of roughly 40:1 to 1000:1, depending
on the assumed duration (and therefore instrument sensitivity) chosen for the
set of bursts as well as the assumed values for the local event rates of both poulations.  For $\rho_{0,LL} = 100 \rm~Gpc^{-3}~yr^{-1}$ (with $\rho_{\rm 0,HL}$ maintained at 1$\rm~Gpc^{-3}~yr^{-1}$), a duration (i.e. $t_{90}$) of 300s gives a ratio of 218:1 HL- to LL-bursts, in general agreement with observation.  If the rate for these bursts is increased to 200 $\rm~Gpc^{-3}~yr^{-1}$ or 400 $\rm~Gpc^{-3}~yr^{-1}$, the durations that give reasonable results drop significantly, to 120 sec and 20 sec, respectively.  This observable ratio is difficult to gauge, however,
due to the few LL-GRBs that actually pass the threshold condition for
instrument sensitivity and the fact that there could be a range of
durations for individual bursts as well as uncertainty in the local rates.  A change in both the duration, $\rho_{0,LL}$, and/or $\rho_{0,HL}$ will modify the set of parameters that will give the correct ratio, as shown above.  For example, if one increases the duration from 120 sec to 500 sec and maintains a rate of 200 $\rm~Gpc^{-3}~yr^{-1}$, the number of LL-bursts detected increases about three-fold, significantly lowering the ratio.  A similar change in ratio will occur when shifting the values of $\rho_0$.  Small changes in the LF parameters of HL bursts (e.g. $L_B = 6.85\times 10^{52}\rm~erg~s^{-1}$ modified to $9.85 \times 10^{52}\rm ~erg~s^{-1}$) do not significantly affect this ratio.
More sensitive detectors
(e.g. JANUS, Roming et al. 2008; EXIST, Grindley 2006) are crucial
for the amassing of LL-burst data and will greatly assist determining typical timescales of these bursts, which affects the sensitivity of the detector, as well as further constrain the relative ratio and in generally improve statistics.  The simulated LL-component should not significantly affect
the bulk of the $L-z$ distributions and the $\log N-\log P$ distribution, but
in the meantime gives rise to the desired LL-GRB events. Due to small number
statistics, the LL-component parameters cannot be well constrained, especially
for $\alpha_{\rm 1,LL}$. In any case, a set of parameters that can best reproduce
the data can be obtained, which are summarized in Table 2.

\begin{figure}
\includegraphics[angle=0,scale=0.9]{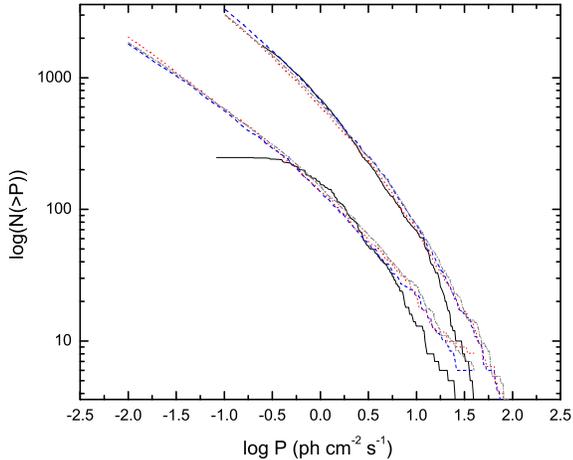}
\caption{Two-component LF model fits to BATSE (top curves) and Swift/BAT (lower curves)
$\log N-\log P$ distributions. The solid (black) curves are the observations,
the dashed (blue) curves are the best fit parameters from the two-dimensional contour, the dotted (red) curves second peak in the probability distribution, and the dash-dot (gray) curves
represent the middle parameters in the maximum of the $p_{KS,t}$ space (see table 3 for details).}
\label{2LF}
\end{figure}

Figures \ref{2LF}-\ref{2D} graphically present the simulated results with the constrained
parameters in Table 2 against the observational data, each set of graphs depicting a different constraint.
Figure \ref{2LF} shows the observed
$\log N-\log P$ distributions
superimposed with the simulated distributions for the two-component
LF model.  In addition to
the BATSE sample (top curves), we also compare with the {Swift}/BAT sample
(lower curves). The observed
distributions seem to level off towards the low photon flux end.
However, this is an effect created by the approach to the
detection threshold of the particular instrument. The simulated
results are not subjugated to such a cutoff,
which serves as a prediction of the model for future observations by more
sensitive detectors such as JANUS and EXIST.  The simulated results are truncated at $0.01\rm~ph~cm^{-2}~s^{-1}$, roughly 20 times the BATSE sensitivity.
The model also slightly overpredicts the number of bursts in the high luminosity end. However, since the number of very high-L GRBs is a small fraction of the total number of GRBs, this excess does not significantly worsen the fit to the data (see also Dai \& Zhang 2005). The discrepancy might be interpreted as due to small number statistics. In general, the two component model reproduces the observed $\log N-\log P$ distribution much better than the other models presented in Fig.1.

\begin{figure}
\includegraphics[angle=0,scale=0.9]{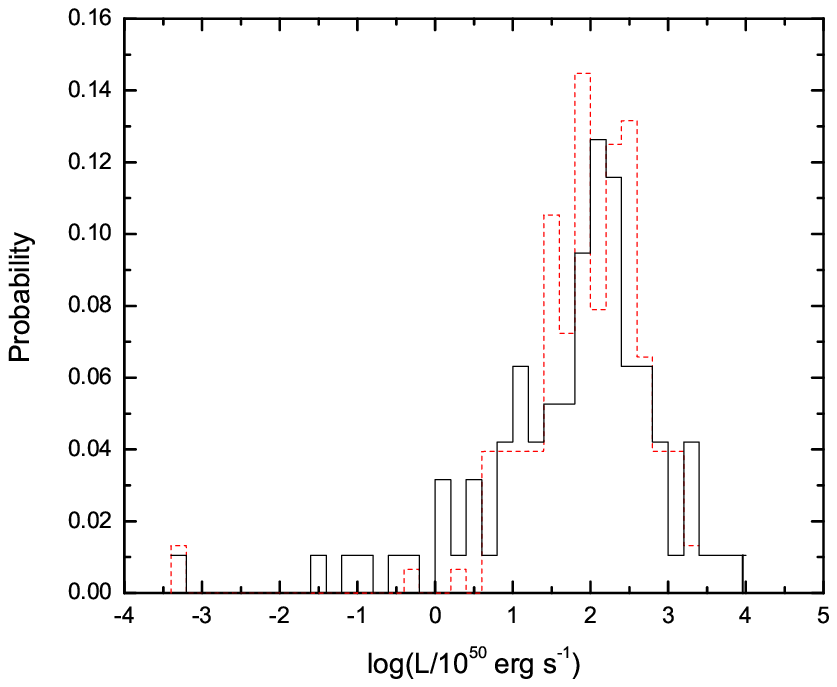}
\includegraphics[angle=0,scale=0.9]{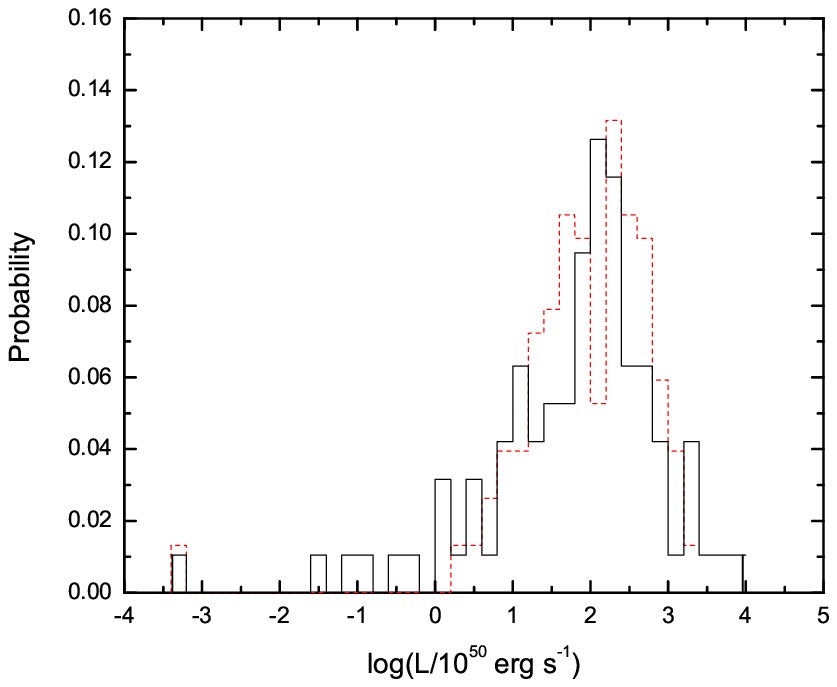}
\includegraphics[angle=0,scale=0.9]{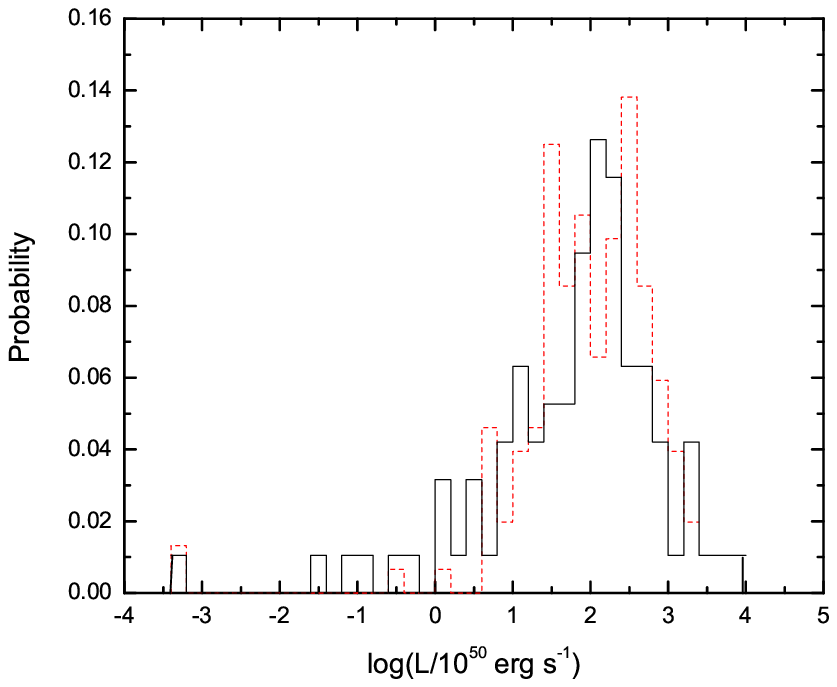}
\caption{1-D luminosity distribution of simulated GRBs (dashed) generated
from the 2-component LF model compared to the observed GRBs (solid).
The panels correspond to best, intermediate, and center K-S probability fits, respectively.
\label{1DL}}
\end{figure}

Next we consider the 1- and 2-D $z$ and $L$ distributions whose results are
summarized in Figs.(\ref{1DL}-\ref{2D}).  The simulated redshift
distribution follows the observed distribution well
through high $z$, except that the observations show a slight overproduction
of bursts with $z > 5$. This may be related to possible evolutionary effects
or additional factors (e.g. metallicity) that determine the GRB rate
(Kister et al. 2007; Daigne et al. 2007; Li 2007; Cen \& Fang 2007), which
we will fully address in a future work.
The simulated 1-D luminosity distribution is also similar
to the observed distribution, broadly peaking at $\sim 10^{52} \rm
~erg~s^{-1}$. A slight deficit of bursts below $\sim
10^{50} \rm ~erg~s^{-1}$ is seen in the data. This effect is most likely caused
by the assumption of the probability for redshift measurement (Eq. \ref{trigger prob}), or
perhaps an effect of a neglected redshift dependence (see
discussion below).
The 2-D $\log L-\log z$ distributions are detailed in Fig.\ref{2D} and
show that the simulated results generally
match the band of observed bursts. There is a small
underproduction of bursts of $10^{53}-10^{54} \rm ~erg~s^{-1}$ at
intermediate redshifts of z$\sim$3 as well as below $\sim 10^{50}
\rm ~erg~s^{-1}$. On the other hand,
any attempt to increase the number of bursts in
this luminosity range would skew the $\log N-\log P$ distribution
at the high photon flux end. A possible cause may be that the fraction of
bursts with redshift measurements in this $(L-z)$ range may be slightly
higher due to the complicated selection effects which are not modeled.
The two simulated LL-bursts are
included in this graph, represented at the lower left hand corner
very near the observed bursts.  The number of HL-bursts plotted reflects the bias in measuring redshifts, namely that only $\sim$ 20 to 30 percent of HL-bursts have a measured redshift.  Low-luminosity bursts are assumed to have a nearly 100 percent redshift detection rate thanks to their proximity.

\begin{figure}
\includegraphics[angle=0,scale=0.9]{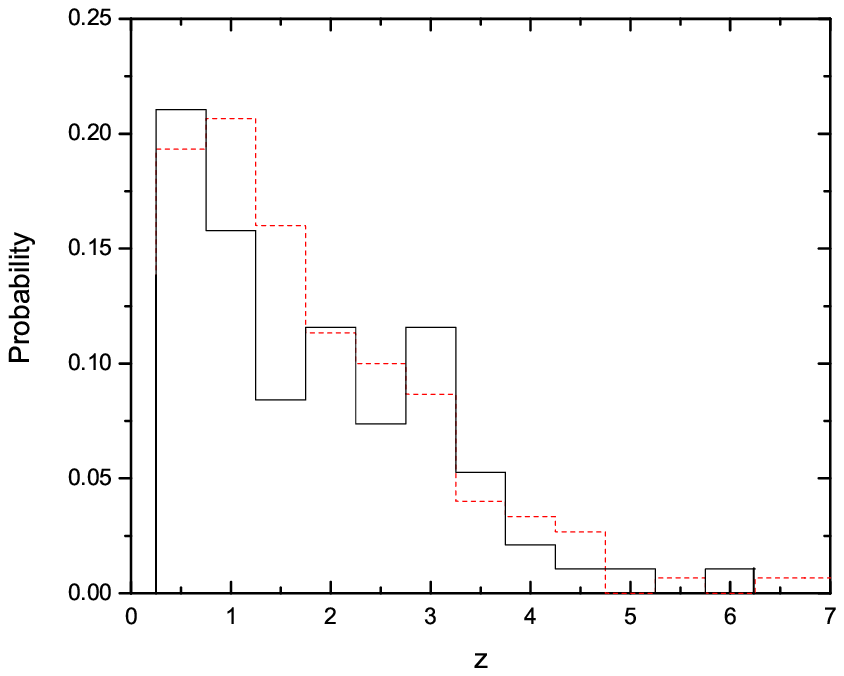}
\includegraphics[angle=0,scale=0.9]{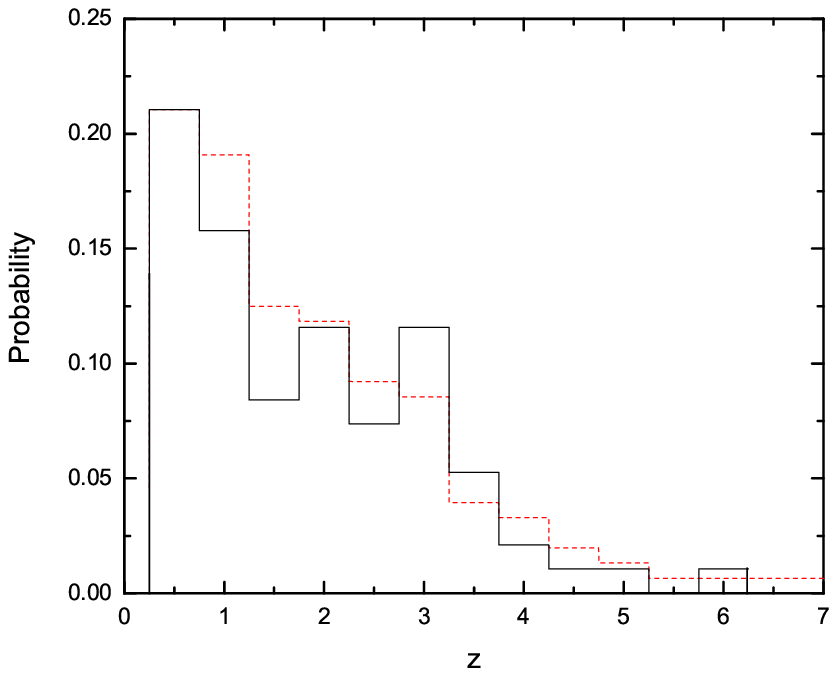}
\includegraphics[angle=0,scale=0.9]{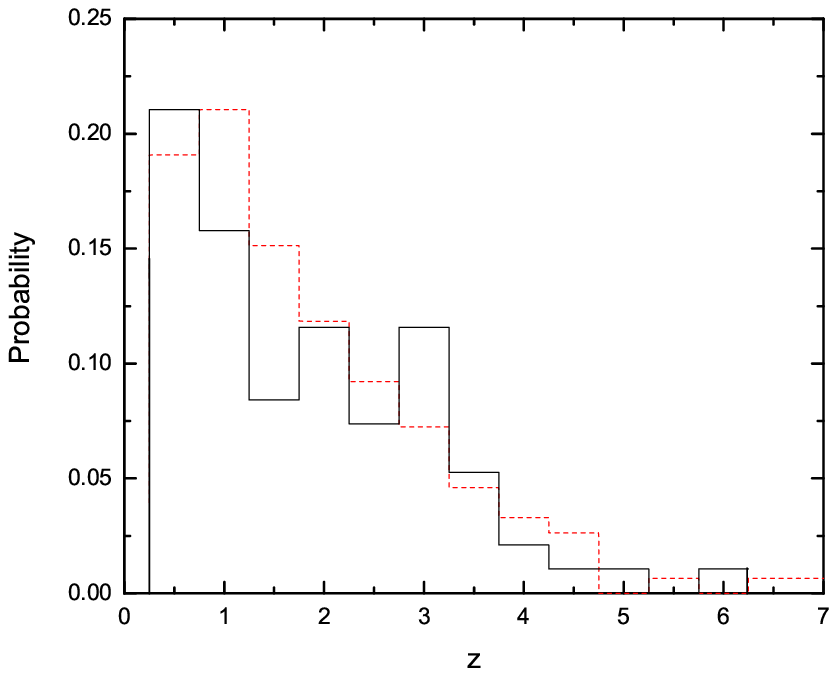}
\caption{1-D redshift distribution of simulated GRBs (dashed) generated
from the 2-component LF model compared to the observed GRBs (solid).
The panels correspond to best, intermediate, and center K-S probability fits, respectively.
\label{1Dz}}
\end{figure}

\begin{figure}\label{2D}
\includegraphics[angle=0,scale=0.9]{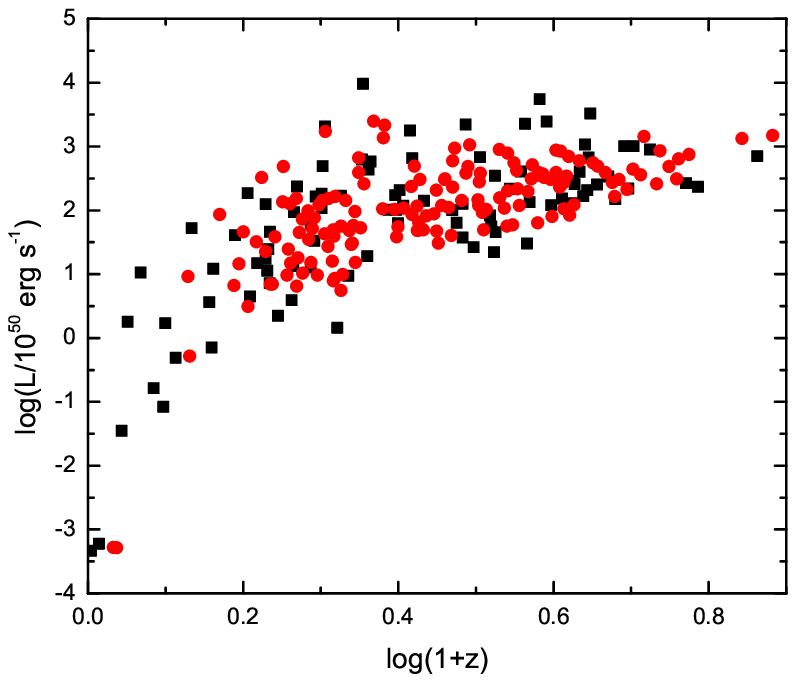}
\includegraphics[angle=0,scale=0.9]{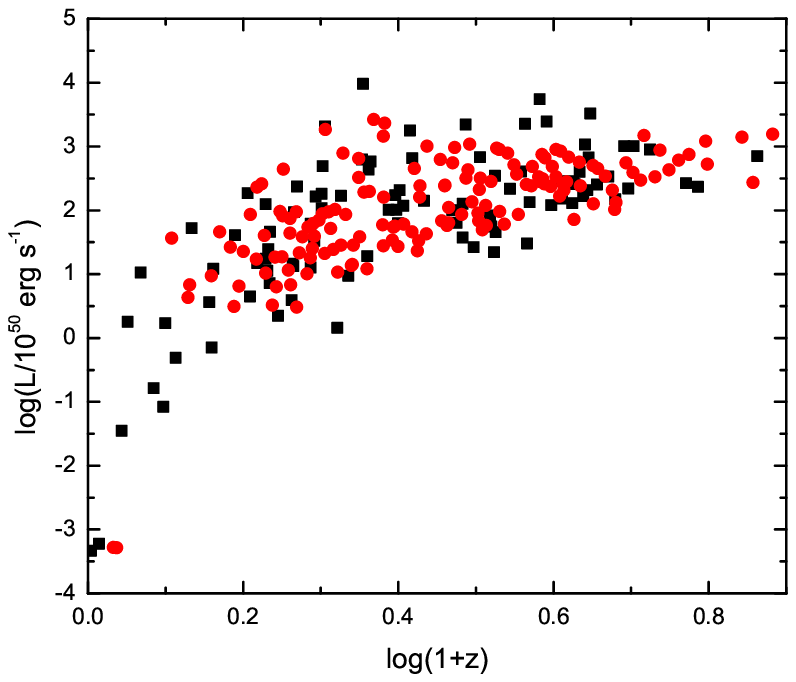}
\includegraphics[angle=0,scale=0.9]{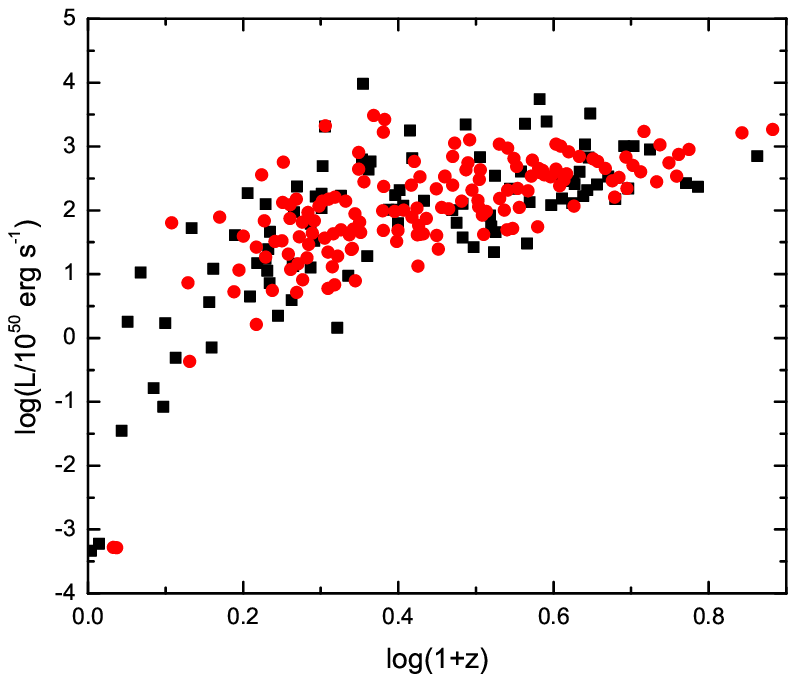}
\caption{2-D $(z-L)$ graph of simulated bursts from the 2-component
LF model (red-circles) as compared to the observed GRBs (black, squares).
The panels correspond to best, intermediate, and center K-S probability fits, respectively.}
\label{2D}
\end{figure}

As mentioned above, intermediate LF parameters were used in producing
Figs. \ref{2LF}-\ref{2D}. Finding more sophisticated functional
forms of the trigger probability, simulating individual burst
timescales, and/or adding terms that evolve with redshift will
most likely increase the overlap of $p_{K-S,z}$ and $p_{K-S,L}$
which together constrain the LF parameters.  Another factor that
influences the K-S probability is the size of the sample of
observed GRBs used in the analysis. As the {Swift} GRB sample
continues to grow, and more bursts are observed with redshift
measurements, the statistical possibilities for analysis will be also
increased. To date, a total of 95 bursts have redshift
measurements, 54 of those coming from {Swift} localizations.

\section{Conclusions and Discussion}

By utilizing Monte Carlo simulations and various test criteria we
are able to further constrain the form and parameters of the LF of
long GRBs. After confronting model results with various observational
criteria, including 1-D and 2-D $L-z$ distributions of the redshift-known
GRBs and the $\log N - \log P$ distributions of both CGRO/BATSE and
Swift/BAT GRBs, we conclude that various one-component LF models
discussed by previous authors (e.g. Guetta et al. 2004, 2005;
Guetta \& Della Valle 2007) are insufficient to account for all the data.  As the luminosity function parameters are modified to accommodate for the observed LL-bursts, these models cause an overproduction of bursts at low and intermediate redshifts that is irreconcilable with the observed distributions.  Although a PL or BPL would seem a simple and straight-forward solution to the LF problem, our reasoning and analysis imply 
that a two-component LF model (Liang et al. 2007, Coward 2005)
is necessary.
The latter model implies an event rate of local LL-GRBs
of $\sim 100-400 \rm~Gpc^{-3}~yr^{-1}$ at an $L_{B,LL}$ of $L\sim 10^{47}~{\rm
erg~s^{-1}}$, which is much larger than that of HL-GRBs
($\sim 1 \rm~Gpc^{-3}~yr^{-1}$).  In addition, as mentioned in Liang et al. (2007), the functional form of the LL-bursts LF is quite uncertain, especially below the break luminosity.  Most constraints are drawn from the more numerous HL observations, although important information about local rate and break luminosities for both distributions can be drawn simply from the number of LL-events detected within the time of Swift's operation.  Other effects that add difficulty to the analysis, but are addressed as fully as possible, include the inhomogeneity of the burst sample, redshift detection selection effects, and the relatively small sample size.

A recent development that may affect the local rate determination is
the serendipitous discovery of  a very low luminosity (peak luminosity
$\sim 6.1 \times
10^{43}~{\rm erg ~s^{-1}}$) X-ray transient, XRF 080109,
by Swift XRT (Soderberg et al. 2008). This event is associated with
SN 2008D. Although it has been suggested that the X-ray emission may
be related to shock breakout (Soderberg et al. 2008), the possibility
that the X-ray emission is the jet emission from a very low luminosity
X-ray flash has been suggested and cannot be ruled out from the data
(Xu et al. 2008; Li 2008). In particular, the non-thermal spectrum
of XRF 080109 makes it different from the thermal X-ray emission
component discovered in XRF 060218, another shock-breakout emission
candidate claimed in the literature (Campana et al. 2006). We
therefore regard the origin of XRF 080109 inconclusive. If it is
indeed a very low luminosity LL-GRB, its very high event rate
(Soderberg et al. 2008; Xu et al. 2008) is consistent with the
conclusion of this paper that LL-GRBs form a distinct new component
in LF, and the event rate increases to even higher values towards
low luminosities. This model also predicts the existence of X-ray
flashes in the luminosity range of $10^{44}-10^{46}~{\rm erg~s^{-1}}$
that bridge XRF 080109 and XRF 060218.

The current LL-GRB sample is too small to address whether they follow
the same empirical correlations as HL-GRBs, such as the lag-luminosity
relation (Norris et al. 2005), variability - luminosity relation
(Reichart et al. 2001),  spectral peak energy - isotropic energy
relation (Amati et al. 2002), etc.
Apparent discrepancy exists for some (e.g. GRB 980425 as an outlier
of the Amati-relation), but consistency exists for others (e.g.
GRB 060218 satisfies the Amati and lag - luminosity relations,
Liang et al. 2006; Amati 2006). On the other hand, these correlations
are closely related to the radiation physics, which
is not directly related to the central engine and progenitor of
the bursts. The GRB fireball picture is very generic. Bursts with
different types of progenitors may share the same emission physics
and hence, similar empirical correlations.
More data are needed to draw firmer conclusions in this direction.

In the analysis
above we show that this 2-component LF model can interpret observations
from both {Swift} and CGRO/BATSE in various criteria.
Although some accommodations have been made to find common ground
within all tests employed, these criteria shed light on the
luminosity function problem and imply that a two-component LF is
necessary.
Various effects that are not considered in
this work may further affect the luminosity and
redshift distribution of the observed bursts.
Changes to the form of the star
formation rate appear often in the literature and are essential to
the basic assumptions of long GRBs as being associated with the
death of massive stars.  Monte Carlo Simulations provide a useful
tool for probing
this effect, either as different functional forms of the SFR
(Porciani and Madau 2001; Rowan-Robinson 1999; Hopkins and
Beacom 2006) or deviations and evolutions with redshift (Kistler
et al. 2007).   Other effects that might affect the distributions
include an evolution of the luminosity function with redshift
(Lloyd-Ronning et al. 2002) or a dependence on cosmic metallicity
(Li, 2007).
These processes might provide solutions to the deficit of
simulated bursts at high-luminosity and high-redshift, since most
of these effects produce a larger rate of bursts at high redshift.
Consequently, this redshift dependence does not affect the
nearby LL-population.  Understanding how and to what extent each
of these processes affects the luminosity and redshift
distributions is a necessary next step in the constraints of the
luminosity function of GRBs, and we plan to explore them in full
in a future work.

\section*{Acknowledgments}
We thank T. Sakamoto and N. Butler for important discussions regarding
BAT trigger procedure and sensitivity threshold, and P. Jakobsson, R.
Chapman for useful communications.
This work is supported by NASA under grants NNG05GB67G,
NNG05GH92G and the Nevada EPSCoR program, and by the President's
Infrastructure Award from UNLV. 
FJV acknowledges NASA's Nevada Space Grant. EWL
acknowledges the National Natural Science Foundation of China under
grant 10463001.

\clearpage

\begin{tabular}{llllllllllll}
\multicolumn{11}{|c|}{Single Component Luminosity Function Models} \\
\hline
Model & Type$^a$ & $\alpha^b$ & $\beta^c$ & L$_B^d$ & L$_1^e$ & L$_2^f$ & $\rho_0$ & $p_{KS,z}$ & $p_{KS,L}$ & $p_{KS,t}$ \\
\hline
G04 & SPL & -0.7 & - & - & 0.5 & 500 & 1.1 & $0.00234$ & 0.00403 & $9.4\times10^{-6}$ \\
G04 (2) & BPL & -0.1 & -0.7 & 0.5 & 0.005 & 500 & 10 & 0.00018 & 0.00022 & $3.96\times10^{-8} $ \\
G05 (P\&M)$^k$ & BPL & -0.1 & -2.0 & 71 & 71$/\Delta_1^h$ & 71$\Delta_2$ & 0.1 & $1.23\times10^{-9}$ & $1.95\times10^{-7}$ & $2.4\times10^{-14}$ \\
G05 (RR) & BPL & -0.1 & -2.0 & 71 & 71$/\Delta_1$ & 71$\Delta_2$ & 0.1 & $8.1\times10^{-12}$ & $2.6\times10^{-9}$ & $2.1\times10^{-20}$\\
G07 & SPL & -1.6 & - & - & 0.5 & 500 & 1.1 & $1.14\times10^{-24}$ & $2.24\times10^{-19}$ & $2.6\times10^{-43}$  \\
G07 (2) & SPL & -1.6 & - & - & 0.005 & 500 & 200 & N/A & N/A & N/A \\
G07 (3) & SPL & -1.6 & - & - & 5$\times10^{-4}$ & 500 & 200-1800$^j$ & N/A & N/A & N/A \\
\hline
\end{tabular}

Notes:  a) SPL = simple power law, BPL = broken power law b) power law index c) For BPL models, power law index after the break luminosity d) break luminosity for broken power law e) lower luminosity cutoff in units of $10^{50}~\rm erg~s^{-1}$ f) high luminosity cutoff in units of $10^{50}~\rm erg~s^{-1}$ g) local GRB rate in units of Gpc$^{-3}$ yr${-1}$ h) $\Delta_1 = 30, \Delta_2=10$, See Guetta et al. (2005), j) Estimation from BATSE data, corrected to 110-1200 Gpc$^{-3}$ yr${-1}$ for BAT constraints (See Guetta 2007) k) Star forming rate model, Porciani and Madau (P\&M) or Rowan-Robinson (RR).

\clearpage

\begin{tabular}{llllllllll}
\multicolumn{9}{|c|}{Two Component Luminosity Function Model Parameters} \\
\hline
$\alpha_1^{LL}$ & $\alpha_2^{LL}$ & L$_B^{LL}$)$^a$ & $\rho_0^{LL}$$^b$ & $\alpha_1^{HL}$ & $\alpha_2^{HL}$ & L$_B^{HL}$ & $\rho_0^{HL}$ & p$_{KS,t}$$^c$ \\
\hline
0.0 & 3.5 & $10^{47}$ & 100 & 0.425 & 2.5 & $5.2\times10^{52}$ & 1 &$ 0.69 $ \\
0.0 & 3.5 & $10^{47}$ & 100 & 0.5 & 2.5 & $8.1\times10^{52}$ & 1 & 0.474 \\
0.0 & 3.5 & $10^{47}$ & 100 & 0.45 & 2.5 & $7.5\times10^{52}$ & 1 & 0.167 \\
\hline
\end{tabular}

Notes: a) $\rm erg~s^{-1}$ b) $\rm Gpc^{-3}~yr^{-1}$ c) Total K-S probability, $p_{KS,t} = p_{KS,L}\times p_{KS,z}$

\clearpage

\begin{tabular}{llll|}
\multicolumn{4}{|c|}{Observed Gamma-Ray Bursts Data} \\
\hline
GRB ID & $z$ & $\log (L/{\rm erg~s^{-1}})$~$^a$ & Reference \\
\hline
970228 & 0.695 & 51.24 & 1\\
970508 & 0.835 & 51.16 & 2,3\\
970828 & 0.9578 & 51.52 & 4\\
971214 & 3.42 & 52.83 & 5\\
980326 & 1.00 & 52.03 & 6\\
980613 & 1.096 & 50.16 & 7\\
980703 & 0.966 & 52.21 & 8\\
990123 & 1.6 & 53.25 & 9\\
990506 & 1.3 & 52.63 & 10\\
990510 & 1.619 & 52.81 & 11\\
990705 & 0.842 & 51.97 & 12\\
990712 & 0.434 & 50.56 & 13\\
991208 & 0.706 & 51.39 & 14\\
991216 & 1.02 & 53.31 & 15\\
000301C & 2.03 & 52.10 & 16\\
000418 & 1.118 & 52.22 & 17\\
000926 & 2.066 & 53.34 & 18\\
010921 & 0.45 & 51.08 & 19\\
011121 & 0.360 & 51.72 & 20\\
011211 & 2.14 & 51.42 & 21\\
020405 & 0.69 & 52.09 & 22\\
020531 & 1.00 & 52.25 & 23\\
020813 & 1.25 & 52.79 & 24\\
020903 & 0.25 & 48.92 & 25\\
021004 & 2.3 & 51.34 & 26\\
021211 & 1.006 & 52.69 & 27,28\\
030226 & 1.98 & 51.80 & 29, 30\\
030323 & 3.372 & 53.03 & 31\\
030328 & 1.52 & 52.31 & 32,33,34\\
030329 & 0.168 & 51.02 & 35\\
030429 & 2.65 & 53.35 & 36\\
040701 & 0.2146 & 49.21 & 37\\
040924 & 0.859 & 52.37 & 38\\
041006 & 0.716 & 51.66 & 39\\
050126 & 1.29 & 51.28 & 40\\
050315 & 1.949 & 52.00 & 41\\
050318 & 1.44 & 52.01 & 42\\
050319 & 3.24 & 52.41 & 43\\
050401 & 2.9 & 53.39 & 44\\
050416 & 0.6535 & 51.17 & 45\\
050505 & 4.3 & 52.95 & 46\\
050525 & 0.606 & 52.27 & 47\\
050603 & 2.821 & 53.74 & 48\\
050724 & 0.258 & 50.23 & 49\\
050730 & 3.97 & 52.34 & 50,51\\
050802 & 1.71 & 52.14 & 52\\
050820 & 2.61 & 52.61 & 53\\
050824 & 0.83 & 50.59 & 54\\
\hline
\end{tabular}

\clearpage

\begin{tabular}{llll|}
\multicolumn{4}{|c|}{Table 3 continued} \\
\hline
GRB ID & $z$ & $\log(L/{\rm~erg~s^{-1}})$ & Reference \\
\hline
050904 & 6.29 & 52.85 & 55\\
050908 & 3.344 & 52.22 & 56,57\\
051016B & 0.9364 & 51.10 & 58\\
051109 & 2.346 & 52.54 & 59\\
051111 & 1.55 & 52.07 & 60\\
051221 & 0.5465 & 51.61 & 61\\
051227 & 0.714 & 50.86 & 62\\
060115 & 3.53 & 52.40 & 63\\
060124 & 2.296 & 51.91 & 64\\
060206 & 4.04998 & 53.01 & 65\\
060210 & 3.91 & 53.01 & 66\\
060614 & 0.125 & 50.25 & 67\\
980425 & 0.0085 & 46.67 & 68\\
060218 & 0.0331 & 46.78 & 69\\
031203 & 0.105 & 48.55 & 70\\
050826 & 0.297 & 49.69 & 71\\
050922C & 2.199, 2.198 & 52.83 & 72, 104\\
060418 & 41.489 & 52.24 & 73,74\\
060502 & 1.51 & 51.80 & 75\\
060510B & 4.9 & 52.43 & 76\\
060512 & 0.4428 & 49.85 & 77\\
060522 & 5.11 & 52.37 & 78\\
060526 & 3.21, 3.221 & 52.46 & 79, 104\\
060604 & 2.68 & 51.48 & 80\\
060605 & 3.78 & 52.18 & 81\\
060607 & 3.082 & 52.46 & 82\\
060707 & 3.425 & 52.19 & 83\\
060714 & 2.71, 2.711 & 52.13 & 84, 104\\
060904B & 0.703 & 51.05 & 85\\
060906 & 3.686 & 52.53 & 86\\
060912 & 0.937 & 51.79 & 87, 105\\
060926 & 3.208 & 52.11 & 88\\
061004 & 0.3 & 52.60 & 106\\
061007 & 1.262 & 54.01 & 90\\
061110 & 0.757 & 50.35 & 91\\
061110B & 3.44 & 53.52 & 92\\
061121 & 1.314 & 52.73 & 93\\
061222B & 3.4 & 52.32 & 94\\
070110 & 2.352 & 51.64 & 95\\
070208 & 1.165 & 50.97 & 96\\
070306 & 1.497 & 52.01 & 97\\
070318 & 0.84 & 51.13 & 98\\
070411 & 2.954 & 52.08 & 99\\
070506 & 2.31 & 51.7661 & 100\\
070529 & 2.4996 & 52.28 & 101\\
070611 & 2.04 & 51.56 & 102\\
070612 & 0.617 & 50.64 & 103\\
\hline
\end{tabular}

\clearpage

References:
1) Djorgovski, S. G., et al., 1997, GCN 289
2) Bloom, J. S., et al., 1997, GNC3 30
3) Zharikov, S.V., et al., 1997, GCN3 31
4) Djorgovski, S. G., et al., 2001b, arXiv:astro-ph/0107539v1
5) Hogg, D. W., Turner, E. L., 1998, GCN 150
6) Fruchter A., et al., 2001a, GCN 1029
7) Djorgovski, S. G., et al., 1999, GCN 189
8) Djorgovski, S. G., et al., 1998 GCN 137
9) Gal, R. R., et al., 1999, GCN 213
10) Bloom, J. S., et al., 2001, arXiv:astro-ph/0102371v1
11) Vreeswijk, P. M., et al., 1999a, GCN 324
12) Le Floc'h et al. 2002, ApJ, 581, L81
13) Galama, T. J., et al., 1999 GCN 388
14) Dodonov, S. N., et al., 1999, GCN 475
15) Vreeswijk, P. M., et al., 1999b GCN 496
16) Castro, S. M., et al., 2000, GCN 605
17) Bloom, J. S., et al., 2000, GCN 661
18) Fynbo, J. P. U., et al, 2000, GCN 807
19) Djorgovski, S. G., et al., 2001a, GCN 1108
20) Infante, L., et al., 2001, GCN 1152
21) Fruchter, A., et al., 2001b, GCN 1200
22) Masseti, N., et al., 2002, GCN 1330
23) Kulkarni, S. R., et al., 2002, GCN 1428
24) Price, P. A., et al, 2002, GCN 1475
25)  Soderberg, A. M., et al., 2002, GCN 1554
26) Chornock, R., et al., 2002, GCN 1605
27) Vreeswijk, P., et al., 2003a, GCN 1785
28) Della Valle, M., et al., 2003, GCN 1809
29) Greiner, J., et al., 2003a, GCN 1886
30) Price, P. A., et al., 2003, GCN 1889
31) Vreeswijk, P., et al., 2003b, GCN 1953
32) Martini, P., et al., 2003, GCN 1980
33) Rol, E., et al., 2003, GCN 1981
34) Maiorano, E., et al., 2006, arXiv:astro-ph/0601293v1
35) Greiner, J., et al., 2003b, GCN 2020
36) Weidinger, M., et al., 2003, GCN 2196
37) Kelson, D. D., 2004, GCN 2627
38) Wiersema, K., et al., 2004, GCN 2800
39) Price, P. A., et al., 2004, GCN 2791
40) Berger, E., et al., 2005b, GCN 3088
41) Kelson, D., Berger, E., 2005, GCN 3101
42) Berger, E., Mulchaey, J., 2005, GCN 3122
43) Fynbo, J. P. U., et al., 2005a, GCN 3136
44) Fynbo, J. P. U., et al., 2005b, GCN 3176
45)  Cenko, S. B., et al., 2005, GCN 3542
46) Berger, E., et al., 2005a, GCN 3368
47) Foley, R. J., et al., 2005a, GCN 3483
48) Berger, E., Becker, G., 2005, GCN 3520
49) Prochaska, J. X., et al., 2005a, GCN 3700
50) Chen, H. W., et al., 2005, GCN 3709
51) DÕElia, V., et al., 2005, GCN 3746
52) Fynbo, J. P. U., et al., 2005c, GCN 3749
53) Prochaska, J. X., et al., 2005b, GCN 3833
54) Fynbo, J. P. U., et al., 2005d, GCN 3874
55) Kawai, N., et al., 2005, GCN 3938
56) Fugazza, D., et al., 2005, GCN 3948
57) Foley, R. J., et al., 2005b, GCN 3949
58) Soderberg, A. M., et al., 2005, GCN 4186
59) Quimby, R., et al., 2005, GCN 4221
60) Hill, G., et al., 2005, GCN 4255
61) Berger, E., Soderberg, A. M., 2005, GCN 4384
62) Foley, J. R., et al., 2005c, GCN 4409
63) Piranomonte, S., et al., 2006, GCN 4520
64) Cenko, S. B., et al., 2006a, GCN 4592
65) Prochaska, J. X., et al., 2006, GCN 4593
66) Cucchiara, A., et al., 2006a, GCN 4729
67) Price, P. A., et al., 2006a, GCN 5275
68) Holland, S., et al, 2000, GCN 704
69) Mirabal, N., Halpern, J. P., 2006, GCN 4792
70) Prochaska, J. X., et al., 2003, GCN 2482
71) Halpern, J. P., Mirabal, N., 2006, GCN 5982
72) Jakobsson, P., et al., 2005, GCN 4017
73) Vreeswijk, P., Jaunsen A., 2006, GCN 4974
74) Dupree, A. K., et al., 2006, GCN 4969
75) Cucchiara, A., et al., 2006b, GCN 5052
76) Price, P. A., 2006b, GCN 5104
77) Bloom, J. S., et al., 2006a, GCN 5217
78) Cenko, S. B., et al., 2006b, GCN 5155
79) Berger, E., Gladders, M., 2006, GCN 5170
80) Castro-Tirado, A., J., et al., 2006, GCN 5218
81) Still, M., et al., 2006, GCN 5226
82) Ledoux, C., et al., 2006, GCN 5237
83) Jakobsson, P., et al., 2006a, GCN 5298
84) Jakobsson, P., et al.., 2006b, GCN 5320
85) Fugazza, D., et al., 2006, GCN 5513
86) Vreeswijk, P., et al., 2006, GCN 5535
87) Jakobsson, P., et al., 2006c, GCN 5617
88) DÕElia, V., et al., 2006, GCN 5637
89) Jakobsson, P., et al., 2006d, GCN 5698
90) Osip, D., et al., 2006, GCN 5715
91)Thoene, C. C., et al., 2006, GCN 5812
92) Fynbo, J. P. U., et al, 2006b, GCN 5809
93) Bloom, J. S., et al., 2006b, GCN 5826
94) Berger, E., 2006, GCN 5962
95) Jaunsen, A. O., et al., 2007a, GCN 6010
96) Cucchiara, A., et al., 2007, GCN 6083
97) Jaunsen, A. O., et al., 2007b, GCN 6202
98) Jaunsen, A., O., et al., 2007c, GCN 6216
99) Jakobsson, P., et al., 2007, GCN 6283
100) Thoene, C. C., et al., 2007a GCN, 6379
101) Berger, E., et al., 2007, GCN 6470
102) Thoene, C. C., et al., 2007b, GCN 6499
103) Cenko, S. B., et al., 2007, GCN 6556
104) Jakobsson, P. et al., 2006e, A\&A, 460, L13
105) Levan, A. J., et al., 2007, MNRAS, 378, 1439
106) Jakobsson, et al., 2006f, GCN 5702

Notes: (a) The spectral parameters and observed fluences in BAT band are
taken from Sakamoto et al. 2007. In order to correct the observed
fluence to the $1-10^4$ keV band in the burst rest frame, we derive
the $E_p$ of each bursts with the relation between $E_p$ and the
spectral power-law index (Zhang et al. 2007; Sakamoto et al. 2007) in
the BAT band and assume that $\Gamma_1=-1$ and $\Gamma_2=-2.3$ for all
bursts.

\clearpage

\bsp

\label{lastpage}

\end{document}